\documentclass[useAMs,a4paper]{mn2e}
\usepackage{savesym}
\usepackage{graphicx}
\expandafter\let\csname equation*\endcsname\relax
  \expandafter\let\csname endequation*\endcsname\relax 
\usepackage{subfig}
\usepackage{amsmath}
\usepackage{amssymb}
\usepackage{verbatim}
\usepackage[yyyymmdd,hhmmss]{datetime}
\usepackage{array}
\usepackage{times}
\usepackage[total={17.8cm,24.0cm},centering]{geometry} 
\usepackage{color}

\newcommand{\be}{\begin{equation}}
\newcommand{\beq}{\begin{equation}}
\newcommand{\ee}{\end{equation}}
\newcommand{\eeq}{\end{equation}}
\newcommand{\eea}{\end{eqnarray}}
\newcommand{\bea}{\begin{eqnarray}}

\title[A maximum TDE X-ray luminosity scale]{A maximum X-ray luminosity scale of disc-dominated tidal destruction events}
\author [Andrew Mummery]{Andrew Mummery\thanks{E-mail:
andrew.mummery@physics.ox.ac.uk}
\\
Oxford Astrophysics, Denys Wilkinson Building, Keble Road, Oxford, OX1 3RH, United Kingdom}
\begin{document}

\date{}

\pagerange{\pageref{firstpage}--\pageref{lastpage}} \pubyear{2021}

\maketitle

\label{firstpage}

\begin{abstract} 
We {develop a} model describing the {dynamical and observed} properties of disc-dominated TDEs {around} black holes {with the lowest masses} ($M \lesssim {\rm few} \times 10^{6} M_\odot$). TDEs around black holes with the lowest masses are most likely to reach super-Eddington luminosities at early times in their evolution.  By assuming that the amount of {stellar debris} which can form into a compact accretion disc is set dynamically by the Eddington luminosity, we make a number of interesting {and}  testable predictions about the observed properties of bright soft-state X-ray TDEs and optically bright, X-ray dim TDEs. We {argue} that TDEs around black holes of the lowest masses will expel the vast majority of their gravitationally bound debris {into a radiatively driven outflow}. A large-mass outflow will obscure the innermost X-ray producing regions, leading to a population of low black hole mass TDEs which are only observed at optical \& UV energies. TDE discs evolving with bolometric luminosities comparable to their Eddington luminosity will have near constant (i.e. black hole mass independent) X-ray luminosities, of order $L_{\rm X, max} \equiv L_M \sim 10^{43} - 10^{44}$ erg/s. The range of luminosity values stems primarily from the range of allowed black hole spins. A similar X-ray luminosity limit exists for X-ray TDEs in the hard ({Compton scattering dominated}) state, and we therefore predict that the X-ray luminosity of the brightest X-ray TDEs will be at the scale $L_M(a) \sim 10^{43}-10^{44}$ erg/s, independent of black hole mass {and accretion state}. {These predictions} are {in strong agreement} with the properties of the existing population ({$\sim 40$ sources}) of observed TDEs. 
\end{abstract}

\begin{keywords}
accretion, accretion discs --- black hole physics --- transients, tidal disruption events
\end{keywords}
\noindent
%Complied at \today\ \currenttime\ .

\section{introduction} 
{Stars which reside in galactic centres can on occasion be perturbed onto orbits which bring them perilously close to their galaxies central black hole.} The tidal destruction and subsequent accretion of {these stars} by the supermassive black hole, a so-called tidal disruption event (TDE), {produces bright transient emission from otherwise quiescent galactic centres.} 

{The  black holes and disrupted stars involved in TDEs are expected to inhabit a broad parameter space. Galactic centre  black holes span} many orders of magnitude in mass, {and are expected to be able to tidally disrupt a solar-mass star from masses $M \sim 10^5 M_\odot$ up to masses of order $M\sim 10^8M_\odot$ (Hills 1975). Yet more rapidly rotating Kerr black holes can tidally disrupt solar-type stars at even higher masses $M \sim 10^9 M_\odot$ (Kesden 2012). In addition, in a sufficiently large population of  TDEs it is likely that the disrupted stars span over an order of magnitude in mass, $M_\star \sim 0.05-1 M_\odot$.  } 

{Given this huge range of potential system parameters, it is perhaps unsurprising that TDEs have been discovered with a wide range of observed properties. In the last two decades TDEs have been discovered and observed at almost every  frequency. This includes hard X-rays (e.g. Cenko et al. 2012), soft X-rays (e.g. Greiner et al. 2000), optical/UV (e.g. Gezari et al. 2008, van Velzen et al. 2020), infrared (e.g. Jiang {\it et al}. 2016, van Velzen et al. 2016b) and radio frequencies (e.g. Alexander et al. 2016). It is likely that the emission observed at high (UV and X-ray) and  low (radio and infrared) photon frequencies have different physical origins.} 

{TDE emission at high photon energies is generally well modelled as having resulted from an accretion disc (except for those TDEs where the X-ray emission is dominated by a relativistic jet e.g. Burrows {\it et al}. 2011).  The X-ray (Mummery \& Balbus 2020a), late-time UV (van Velzen et al. 2019, Mummery \& Balbus 2020a, b) and late-time optical (Mummery \& Balbus 2020b) light curves of a number of TDEs are all well described by evolving thin-disc models. Slim accretion discs (those with finite aspect ratios) have been invoked to model the evolving X-ray spectra of some TDEs (Wen et al. 2020),  as have sequences of steady-state AGN-type models (e.g. Saxton et al. 2019).   }

{At lower photon energies the physical mechanism invoked to explain the observed emission is often an outflow launched in the early stages of the TDE (Dai {\it et al}. 2018, Metzger \& Stone  2016, Strubbe \& Quataert 2009). These outflow models are expected to explain the properties of some TDE radio outbursts (with other radio bright sources being associated with a jet),  and the optical emission observed from TDEs at early times.  The general idea of these models is that some fraction of the stellar debris produced in the aftermath of a TDE is expelled from the black hole-disc system, this material then reprocesses disc emission to lower observed energies. The debris resulting from the tidal destruction of a star will be gravitationally bound to its central black hole to various different degrees, and so it is not surprising that some material may settle into a disc while other debris will be ejected from the system entirely. X-ray and UV emission stemming from the accretion disc passes through this outflowing material on the way to the observer, where it is absorbed and re-emitted at lower photon frequencies.  }

{Models of outflows from TDEs therefore generally act to suppress observations of high energy (X-ray) TDE emission and to increase the amount of observed optical TDE emission. In fact Dai {\it et al}. (2018) have argued that the properties of all observed TDEs can be unified in a model of this kind. Dai {\it et al.} argued (based on a single GRMHD simulation) that every TDE results in a geometrically thick accretion flow accreting at super-Eddington rates. Outflows launched from this super-Eddington flow can, for certain viewing angles, obscure the inner-most X-ray producing region, leading to detections of only optical/UV radiation. Conversely, for different viewing angles the innermost regions will not be obscured, and bright X-ray emission will be detected. 

The GRMHD simulation which Dai et al. (2018) based their argument upon involved a relatively low mass $M = 5\times10^6M_\odot$, and relatively rapidly rotating $a/r_g = 0.8$ black hole.  We will demonstrate in this paper that TDEs around low mass and rapidly rotating black holes are exactly those TDE systems which are most likely to result in initial super-Eddington accretion rates. However, TDEs occurring around black holes of larger masses will almost certainly form with sub-Eddington accretion discs (e.g. Mummery \& Balbus 2020b), it therefore seems unlikely that the model of Dai {\it et al}. is truly universal.

In this paper we aim to build upon and extended these outflow models in a number of ways.  Firstly, we examine over which regions of TDE (black hole and disc) parameter space outflows are likely to be launched from a TDE disc. We do this by coupling the launching of a radiatively driven outflow to the amplitude of the discs bolometric luminosity in Eddington units. This differs from previous treatments (Metzger \& Stone 2016) which link the launching of an outflow to the rate at which stellar debris returns to the pericentre of the disrupted stars orbit. By coupling the radiatively driven outflow to the properties of the disc luminosity (as opposed to a proxy for the disc luminosity) we are able to quantitatively predict exactly which TDEs will be observationally affected by outflows suppressing their X-ray luminosity. 

Explicitly, we demonstrate that radiatively driven outflows which are coupled to the discs bolometric luminosity will be preferentially launched from black holes of masses lower than a typical mass scale which we call $M_{\rm edd}$. We show that for typical TDE disc parameters  $M_{\rm edd} \simeq 5 \times 10^6 M_\odot$. Although this value depends weakly on both the disc parameters and the black hole spin (eq. \ref{eddmass}), it is typically $M_{\rm edd} \lesssim {\rm few} \times 10^6 M_\odot$.  If these radiatively driven outflows do indeed act to suppress observations of high energy emission, then we would expect there to exist a population of TDEs observed only at UV and optical wavelengths around black holes of the lowest masses. We show in this paper that the existing population of observed TDEs are consistent with this picture. 

In this paper we further extend existing outflow models by examining what effects outflows coupled to the discs bolometric luminosity have on the X-ray luminosity distribution of observed TDEs. We show, analytically and numerically, that TDE discs evolving with bolometric luminosities comparable to their Eddington luminosities  have near constant (i.e. black hole mass independent) X-ray luminosities, of order $L_{\rm X, max} \equiv L_M \sim 10^{43} - 10^{44}$ erg/s. The range of luminosity values here stems primarily from the range of allowed black hole spins. Physically, this near constant X-ray luminosity results  from the competing black hole mass scalings of the peak disc temperature and emitting disc surface area. When a discs bolometric luminosity is comparable to its Eddington luminosity, hotter, lower-area discs produce comparable X-ray fluxes to cooler, larger-area discs. 

If TDE discs do indeed launch radiatively driven outflows when their bolometric luminosities become comparable to their Eddington luminosity then we would expect to see this luminosity scale $L_M \sim 10^{44}$ erg/s imprinted on the TDE X-ray luminosity population, $L_{X, {\rm TDE}} \lesssim L_M$.  In the final sections of this paper we demonstrate that of 24 currently known X-ray bright TDEs no TDE has been observed with an X-ray luminosity brighter than the theoretical limiting scale derived in this paper.

This paper represents the third part of a four paper TDE unification scheme (Mummery \& Balbus 2021a, b, Mummery 2021b). The layout of this paper is as follows: in section \ref{discedd} we discuss the observational justification for an outflow model coupled to high disc luminosities, and formally introduce our model. In section \ref{numres} we numerically analyse the results of a disc luminosity-outflow coupling on the observed X-ray luminosity of TDEs.  In section \ref{obscuresec} we discuss implications of our model for the observed mass distribution of different spectral classes of TDEs, demonstrating that the existing TDE population is in good agreement with our predictions. In section \ref{hardnum} we analyse the maximum X-ray luminosity properties of hard-state TDEs, before analysing the properties of the existing population of X-ray bright TDEs in section \ref{population}.  We conclude in section \ref{conc}. 
}

\section{Disc Eddingtonization \& Outflows }\label{discedd}
{In this section we justify and then formally introduce our disc-outflow model for high-luminosity TDEs.  We begin by outlining the evidence, obtained from X-ray spectral observations, which imply that no TDE has yet been observed in a super-Eddington accretion state. We then recap arguments  for why high-luminosity TDEs will likely produce a radiatively driven outflow. In the second half of this section we introduce our disc outflow model.    }

\subsection{Observational justification} 
In {Mummery \& Balbus 2021a (hereafter Paper I}) we demonstrated that the peak temperature within a time-dependent relativistic accretion disc has an extremely sensitive dependence on the central black hole mass.  Expressed in terms of the disc mass $M_d$, disc $\alpha$-parameter $\alpha$, and black hole mass $M$, the temperature scales in the following way 
\beq\label{temp}
T_p \propto \alpha^{1/3} M_d^{5/12} M^{-7/6} .
\eeq
This strong system parameter dependence then leads to Eddington ratios with the following dependence (Paper I)
\beq\label{edrat}
l \equiv {L_{\rm bol, peak} \over L_{\rm edd}} \propto {\alpha^{4/3} M_d^{5/3} \over M^{11/3}} .
\eeq
Equation \ref{edrat} demonstrates that the peak Eddington ratio of a time dependent accretion disc is an extremely sensitive function of the central black hole mass.  For any given disc mass and $\alpha$ parameter there will exist a limiting black hole mass (which we shall call $M_{\rm edd}$), where $L_{\rm bol, peak} = L_{\rm edd}$. It is then clear from eq. \ref{edrat} that all TDEs with identical disc parameters ($M_d$ and $\alpha$) around black holes with masses lower than this value would produce initially super-Eddington bolometric luminosities. As an important reference value $M_{\rm edd}$ may be determined numerically, and for a Schwarzschild black hole with disc mass $M_d = 0.5M_\odot$ and $\alpha = 0.1$, $M_{\rm edd} \simeq 5 \times 10^6 M_\odot$ (Paper I). 

If TDEs around black holes with masses $M < M_{\rm edd}$ did indeed form super-Eddington accretion flows at early times, then we would expect there to be an abundance of low black hole mass TDEs with X-ray spectra similar to those observed in ultra-luminous X-ray sources (ULXs). ULX spectra are characterised by a much higher comptonization optical depth ($\tau \sim 5-20$), compared to those seen in the X-ray binary (XRB) high-soft state ($\tau < 1$) (Gladstone {\it et al}. 2009). The XRB high-soft state is expected to be the spectral state most similar to TDE discs which form with Eddington ratios $0.01 \lesssim l \lesssim 1$. 

However, there is as of yet no observational evidence for this super-Eddington `ULX-like' accretion state in low black hole mass TDEs (Jonker {\it et al}. 2020). On the contrary, early-time soft X-ray detections of TDE candidates generally find quasi-thermal spectra that are analogous to an XRBs high-luminosity, spectrally soft state (Komossa \& Greiner 1999; Greiner {\it et al}. 2000), particularly in TDEs with good quality early time X-ray spectra (e.g. Saxton {\it et al}. 2012, Miller {\it et al}. 2015, Lin {\it et al}. 2015, Holoien {\it et al}. 2016a, Gezari {\it et al}. 2017, Wevers {\it et al}. 2019b, Jonker {\it et al}. 2020, Wen {\it et al}. 2020). 

Taken at face value this is a rather surprising result. The not unreasonable set of disc parameters $M_d = 0.5 M_\odot$ \& $\alpha = 0.1$ around a Schwarzschild black hole with mass $M = 2\times10^6M_\odot$ would lead to an Eddington ratio at peak of $l \simeq 30$ (or even higher for a rapidly spinning Kerr black hole). We suggest that this observational fact cannot be merely a coincidence of system parameters. 

Rather, we suppose that these observations may be indicating a fundamental property of dynamical disc formation in the aftermath of a TDE. {In this paper we} assume that X-ray TDEs with ULX-like spectra have not been observed because the amount of {stellar debris} that can {form into an accretion disc at radii} close to the central black hole, and are thus hot enough to produce thermal X-rays, is set dynamically by the Eddington limit. 

{There are a number of reasons to suspect that a large fraction of the stellar debris in a TDE may be unable to form into a disc when that disc is producing large bolometric luminosities.} The matter within a TDE disc is gradually deposited, with debris returning to the pericentre of the disrupted stars orbit approximately according to $\dot M_{\rm fb} \sim t^{-5/3}$ (Rees 1988). {This means that if disc luminosities reach large values at early times in the TDE evolution, a large fraction of the stellar debris will still not have returned to the disc. This remaining material is then susceptible to being expelled due to large radiation pressures}. Moreover, much of this returning matter is likely to be only tenuously gravitationally bound (Metzger \& Stone 2016). {In fact,} it is known that not all of the debris mass forms into a disc: exponentially declining light curve components are observed across optical and UV bands at early times, prior to these light curves transitioning to a disc dominated state (Van Velzen et al. 2019, Mummery \& Balbus 2020a, b). We propose that the amount of matter that {forms into an accretion disc} close to the black holes ISCO is set so that, at peak, $L_{\rm bol} \leq L_{\rm edd}$. 

The physical mechanism by which this disc formation process is likely to be mediated is the launching of radiation driven outflows. Models of the effects of  outflows on observed TDE properties already exist (e.g., Metzger \& Stone 2016, Strubbe \& Quataert 2009, Dai {\it et al}. 2018). However, our model differs in one important aspect: we link the launching of the radiative outflow to the discs bolometric luminosity $L$, as opposed to the rate at which debris return to pericentre $\dot M_{\rm fb}$. {In the following section we demonstrate that this coupling} means that radiatively driven outflows will be {dynamically and observationally} important at systematically lower black hole mass scales when compared to these previous works.

\subsection{Analytical model}
{In this work w}e will assume that the amount of material which forms into a compact accretion disc near to the central black hole in the aftermath of a TDE is set so that the peak bolometric luminosity of the resulting accretion disc is equal  to, or less than, the Eddington luminosity. {We call this behaviour ``disc Eddingtonization''.} The remainder of the stellar debris, if any,  is assumed to be ejected from the accretion disc by a radiatively driven outflow. 

We define $M_{\rm deb}$ as the amount of stellar material which remains gravitationally bound to the central black hole after disruption, and which could potentially form into a disc. We further define $l_p$ as the peak Eddington ratio which would occur had {\it all} of  $M_{\rm deb}$ formed into a disc. With these definitions it follows from equation \ref{edrat} that if the surviving disc mass $M_d$ is given by 
\beq\label{survive}
M_d = 
\begin{cases}
 M_{\rm deb}, \quad &l_p < 1, \\ \\
 M_{\rm deb}\,  l_p^{-3/5},  \quad &l_p \geq 1, \\
\end{cases}
\eeq
then the peak Eddington ratio of a disc which forms with mass $M_d$ is $l \leq 1$, with equality in the radiatively driven outflow regime. We assume that the remaining matter is ejected as an outflow $M_{\rm out} = M_{\rm deb} - M_d$:
\beq
M_{\rm out} = M_{\rm deb} \, \left[ 1 - l_p^{-3/5} \right], \quad l_p \geq 1.  
\eeq
As the Eddington ratio is a  function of black hole mass (eq. \ref{edrat}), this can be equivalently written in terms of the ratio between the black hole mass $M$ and the black hole mass at which all of the debris mass could form into a disc with Eddington ratio $l = 1$ ($M_{\rm edd}$), 
\beq
M_{\rm out} = M_{\rm deb} \, \left[ 1 - \left( {M \over M_{\rm edd}}\right)^{11/5} \right], \quad M \leq M_{\rm edd}.  
\eeq
{The critical Eddington mass value $M_{\rm edd}$ is a function of disc parameters, with a scaling which can be found from equation \ref{edrat}, and an amplitude which can be determined numerically (Paper I). The following result is valid for a Schwarzschild black hole:
\beq\label{eddmass}
M_{\rm edd} \simeq 5\times10^6 M_\odot \left({M_d \over 0.5M_\odot}\right)^{5/11} \left({\alpha \over 0.1}\right)^{4/11} .
\eeq
 The amplitude in this relationship remains a function of black hole spin, with higher black hole spins resulting in higher Eddington masses. The Schwarzschild value quoted above is however a good ball-park figure.  }

Potential observational implications resulting from obscuration of the inner disc regions by this radiatively expelled matter will be discussed further in section \ref{obscuresec}. 

An important implication of the surviving disc mass being set so that the peak Eddington ratio is unity is that the peak disc temperature now only depends on the black hole mass, in the following way 
\beq\label{temp2}
T_p \propto M^{-1/4},\quad M \leq M_{\rm edd}.
\eeq
This {well known result} can be verified by combining equations \ref{temp}, \ref{edrat} \& \ref{survive}.  A change in parameter dependence of the peak temperature from an extremely sensitive function of disc and black hole parameters (eq. \ref{temp}), to a weak dependence on only the black hole mass (eq. \ref{temp2}), results in markedly different properties of the X-ray luminosity in the two regimes.  In paper I we demonstrated that the peak X-ray flux from a thermal TDE disc scales, to leading order, like 
\beq\label{flux}
F_X \propto \left({M \over D}\right)^2 \left({k_B T_p \over E_l}\right)^A \exp\left(-{E_l \over k_B T_p} \right),
\eeq
where $E_l$ is the lower energy of the X-ray telescopes band pass, $D$ is the source-observer distance, and the index $A$ depends on assumptions about the ISCO stress and observer orientation angle. For a face-on finite ISCO stress disc $A=2$, whereas for a face-on vanishing ISCO stress disc $A = 3/2$. The different parameter dependences of the peak disc temperature filter through to extremely different X-ray flux dependences (here showing the result for a vanishing ISCO stress):
\beq\label{fluxM}
F_X \propto 
\begin{cases}
 M^{1/4} \exp\left(-C_1 M^{7/6} \right), \quad M > M_{\rm edd}, \\ \\
M^{13/8} \exp\left(-C_2 M^{1/4} \right), \quad M \leq M_{\rm edd}.
\end{cases}
\eeq
In this expression $C_1$ and $C_2$ are dimensional parameters which depend on the disc properties. These two expressions have qualitatively different properties. When sub-Eddington ($M > M_{\rm edd}$) the X-ray flux is given by the product of a polynomial which is a weak function of mass, and a negative exponential which is strongly mass dependent. However, when in the outflow regime ($M < M_{\rm edd}$), the X-ray flux is given by the product of a polynomial which is a  strongly increasing function of mass, and an exponential which is only weakly mass dependent. 

We shall demonstrate numerically that the dual effects of this growing power-law and weakly decaying exponential effectively cancel out, leading to near-constant (i.e. black hole mass independent) X-ray luminosities for TDEs in this regime. 

%%%%%%%%%%%%%%%%%%%%%%%%%%%%%%%%%%%%%%%%

\begin{figure}
  \includegraphics[width=.5\textwidth]{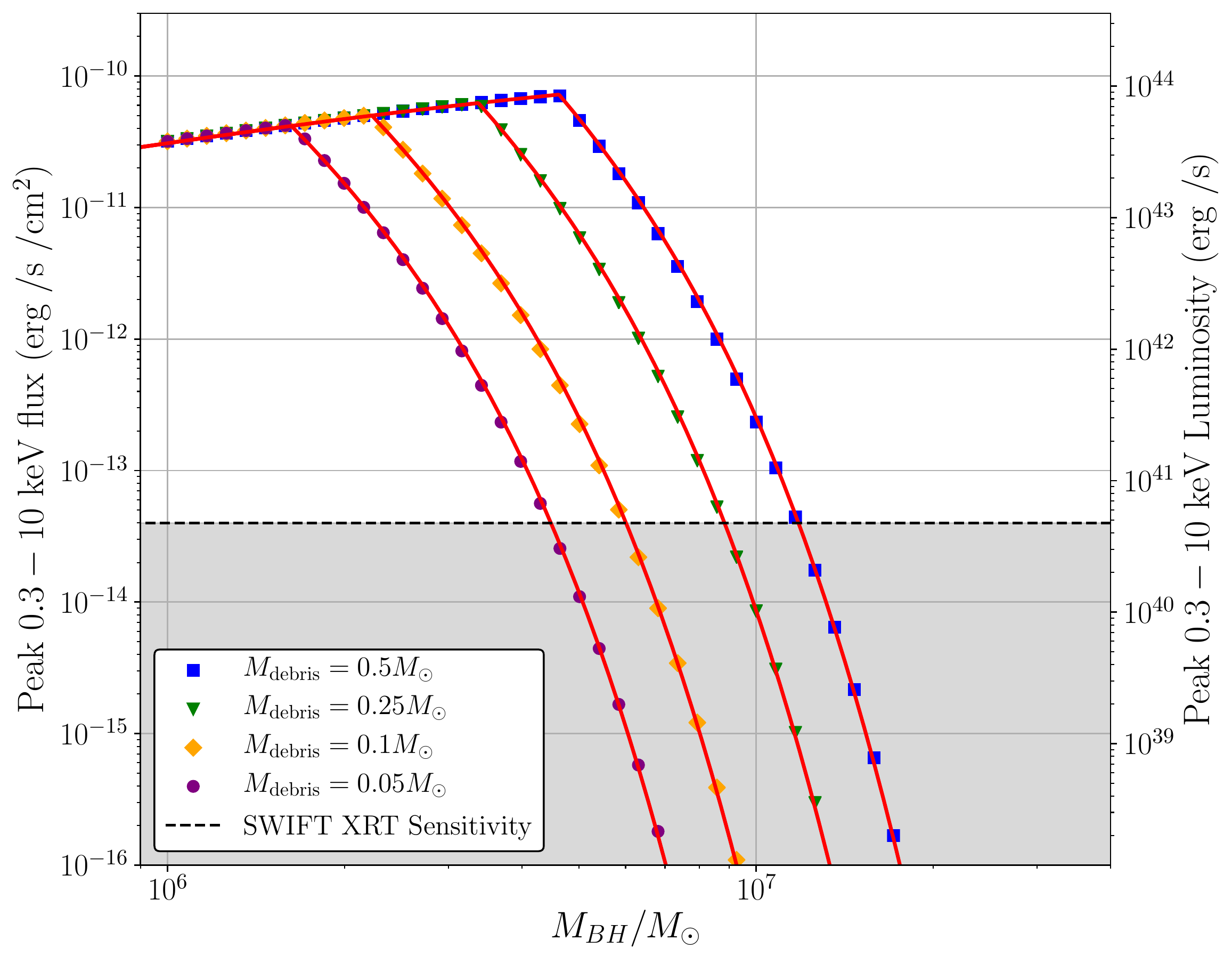} 
 \caption{The peak $0.3$--$10$ keV X-ray flux, as observed for a face-on disc orientation at $100$Mpc, for discs evolving with  $\alpha=0.1$ and a number of different debris masses $M_{\rm deb}$ denoted on plot.  The initial disc mass is equal to $M_{\rm deb}$ except for black hole masses where the bolometric luminosity is found to be super-Eddington. For these black holes the disc mass is reduced until the peak luminosity is equal to the Eddington luminosity. The solid curves are the analytical model of Paper I, with disc mass given by equation \ref{survive}. The near constant X-ray luminosity in the $M < M_{\rm edd}$ regime is independent of the disc microphysics. } 
 \label{simple}
\end{figure}

\section{Numerical results}\label{numres}

\begin{figure}
  \includegraphics[width=.5\textwidth]{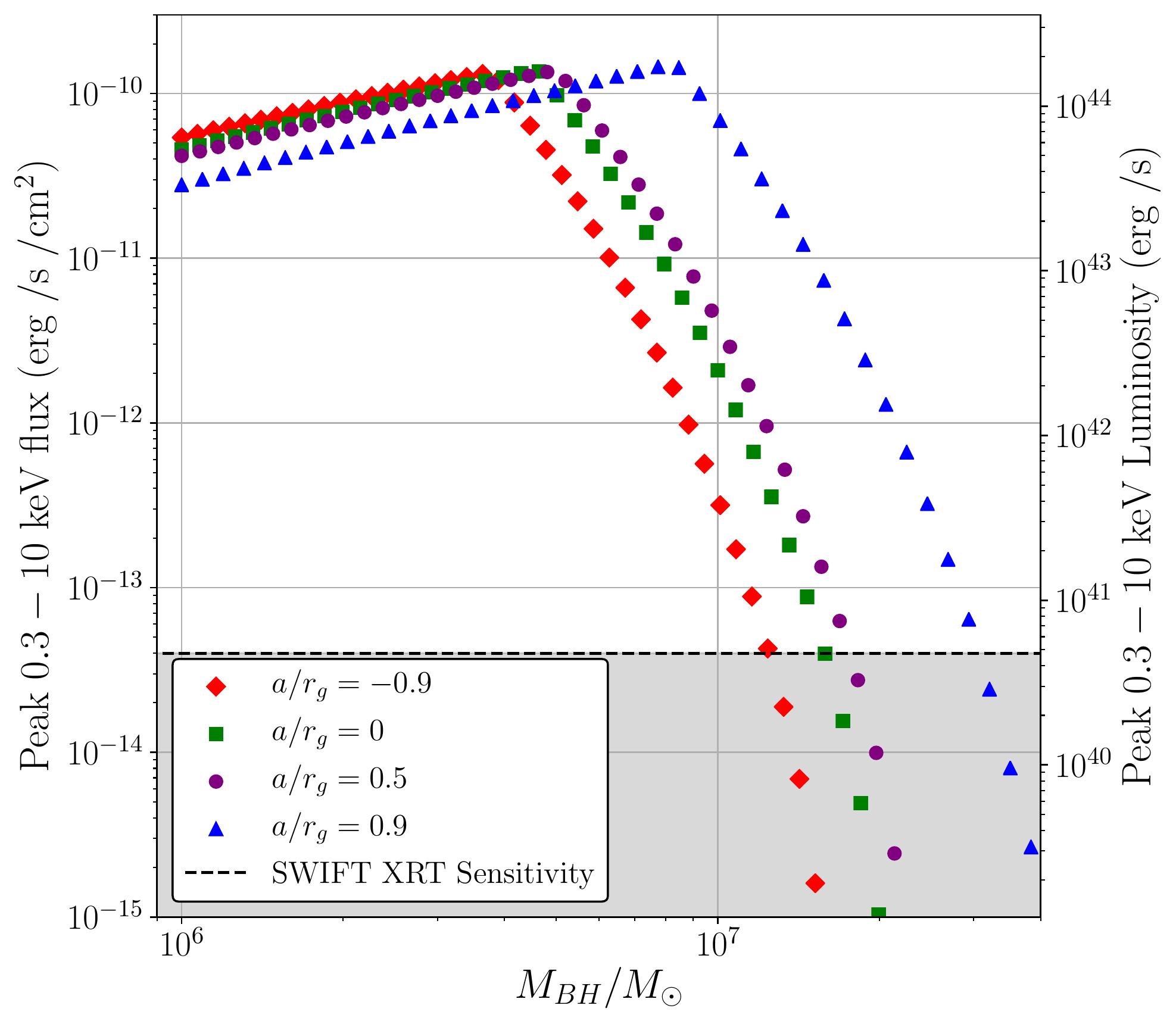} 
 \caption{The peak $0.3$--$10$ keV X-ray flux, as observed at $100$Mpc, for discs evolving with  $\alpha=0.1$ and $M_{\rm deb} = 0.5 M_\odot$. The flux is computed as in figure \ref{simple}, but now with an inclination $\theta_{\rm obs} = 60^\circ$ and a number of different black hole spins, denoted on plot. This figure was produced for discs with a finite ISCO stress. The upper X-ray luminosity scale is only weakly dependent on black hole spin.   } 
 \label{spins}
\end{figure}

In this section we examine the properties of the X-ray luminosity of TDE disc systems in the radiatively driven outflow regime. We assume that the amount of matter which forms into a compact accretion disc in the aftermath of a TDE is set such that the bolometric luminosity of the resulting disc is no brighter than the Eddington luminosity of the central black hole. 

{We shall examine the properties of the discs X-ray luminosity for a wide range of both ISCO stress vales and disc-observer orientation angles, along with other disc parameters. In reality it is unlikely that TDE discs spanning the full range of both of these parameters will actually be observed.  For example, at high luminosity values TDE accretion discs will likely have moderate aspect-ratios (becoming `slim' discs), meaning that discs observed at large disc-observer orientation angles will have their inner-most regions obscured from view by their outer regions. This will act to suppress the  observed X-ray emission from these sources. Furthermore, it is unlikely that high luminosity TDE discs will span a large range of ISCO stress values. We examine the full range of parameter space for these parameters simply so that we are certain that the maximum X-ray luminosity  of TDE discs is robust to variations in all the system parameters.   }

To compute the evolving X-ray luminosity of the TDE disc system we numerically solve the general relativistic thin disc evolution equation (Balbus 2017, Mummery \& Balbus 2019a, b), assuming an $\alpha$ model of the disc turbulence (see Paper I, section 4 for full details).  The disc is initially parameterised by a debris mass $M_{\rm deb}$, and $\alpha$-parameter $\alpha$. We compute the evolving bolometric and X-ray light curves of the resulting disc system. If for a given debris mass $M_{\rm deb}$ and black hole mass $M$ the discs bolometric luminosity exceeds the black holes Eddington luminosity,  the initial disc mass is reduced until the peak of the bolometric light curve equals the Eddington luminosity. For that new disc mass the peak X-ray flux (as observed at a distance of $D = 100$ Mpc) is computed. This method includes the temperature-dependent colour-correction factor $f_{\rm col}(T)$ of Done {\it et al}. (2012), and a fully general relativistic ray-tracing calculation (Appendix A, Mummery \& Balbus 2020a). 

The peak observed $0.3-10$ keV flux calculated as described above, assuming a face-on orientation, a Schwarzschild black hole and a number of different initial debris masses, is displayed in figure \ref{simple}. The red solid curves are the analytical model of Paper I (eq. 46, Paper I), with disc mass given by equation \ref{survive}. 

It is clear to see that the assumption of disc Eddingtonization leads to a maximum observed X-ray luminosity (which we shall denote $L_{M}$)  of TDE disc systems. For the system parameters considered in figure \ref{simple} this is of order $L_{M} \equiv 4\pi D^2 F_{X, \rm max} \simeq 5 \times 10^{43}$ erg/s. The X-ray luminosity scale is independent of the original disc parameters $M_{\rm deb}$ \& $\alpha$, as in this regime the peak disc temperature only depends on the black hole mass (eq. \ref{temp2}), and the X-ray luminosity of TDE disc systems only depends on the hottest temperature within the accretion disc (eq. \ref{flux}). 

\begin{figure}
  \includegraphics[width=.5\textwidth]{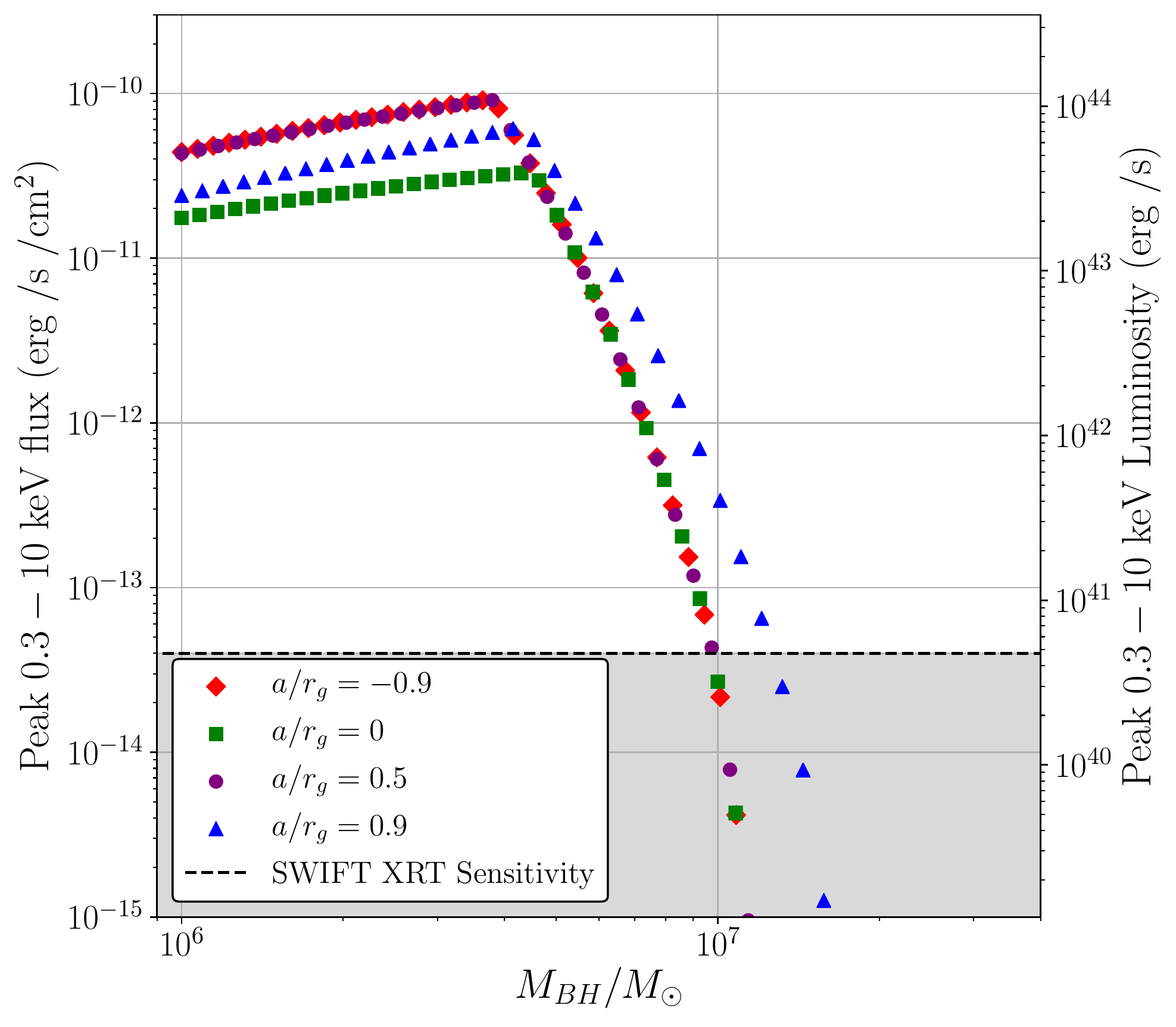} 
 \caption{The peak $0.3$--$10$ keV X-ray flux, as observed at $100$Mpc, for discs evolving with  $\alpha=0.1$ and $M_{\rm deb} = 0.5 M_\odot$. The flux is computed as in figure \ref{spins}, except for the ISCO stress which is now assumed to vanish. The upper X-ray luminosity scale is only weakly dependent on assumptions about the ISCO stress. } 
 \label{VSspins}
\end{figure}
\begin{figure}
  \includegraphics[width=.5\textwidth]{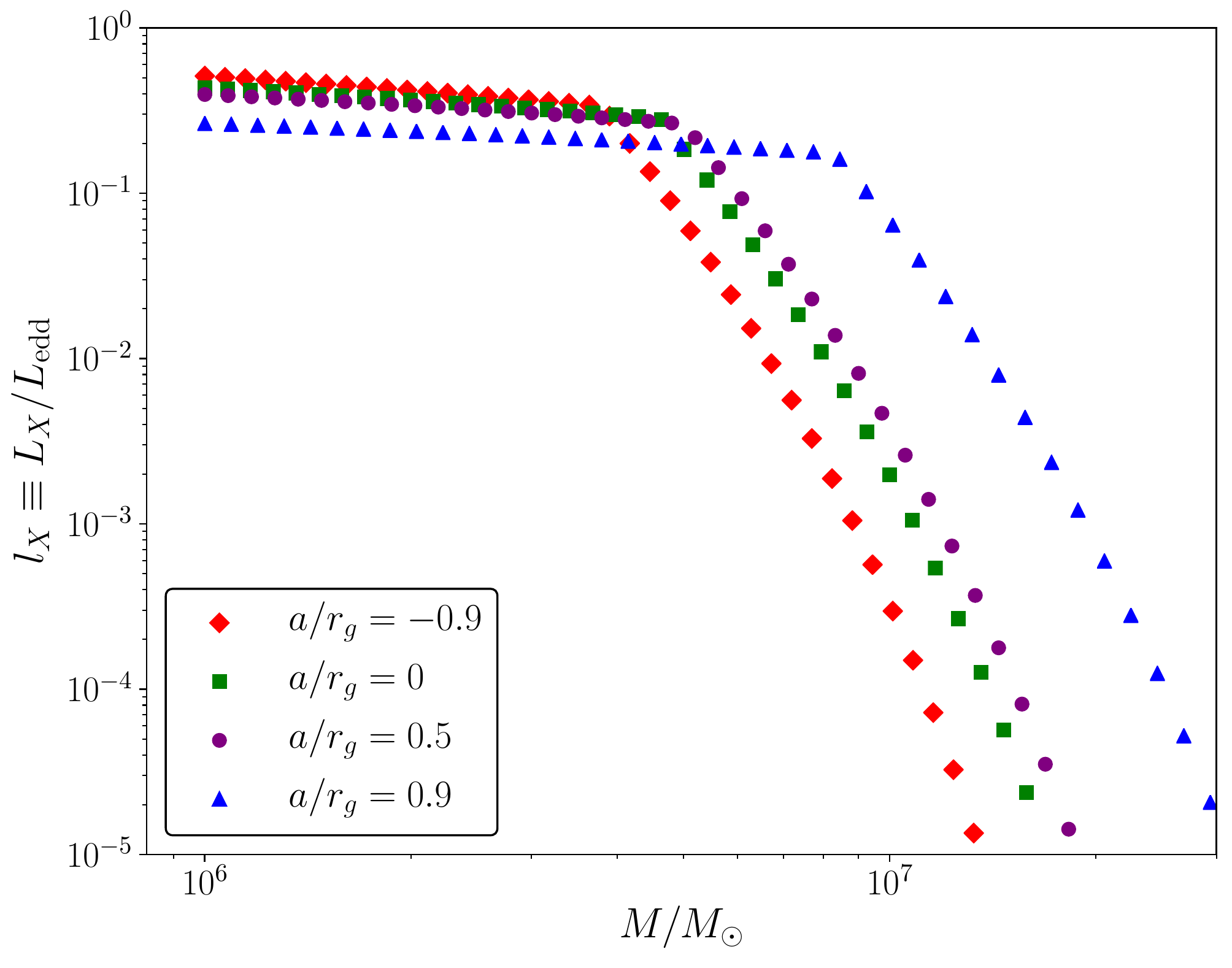} 
 \caption{The peak observed X-ray Eddington ratios of the evolving disc solutions from figure \ref{spins}. Disc models of TDE X-ray evolution predict X-ray Eddington ratios at most $L_X/L_{\rm Edd} \sim {\rm few} \, \times   10^{-1}$, but which can vary by many orders of magnitude over a small range of black hole masses.    } 
 \label{edrats}
\end{figure}

This X-ray luminosity scale is only weakly dependent on black hole spin (figure \ref{spins}), assumptions about the ISCO stress (figure \ref{VSspins}), and orientation angle. The peak X-ray luminosity varyies from $L_M \sim 10^{43} - 10^{44}$ erg/s  depending on the particular spin, ISCO stress and orientation angle chosen.  We therefore expect to observe the imprint of this X-ray luminosity scale on the observed X-ray TDE population. If X-ray bright TDE discs typically form with Eddington ratios close to unity, then the majority of thermal X-ray TDEs should have peak luminosities of order $\sim 10^{43} - 10^{44}$ erg/s. The observational evidence for this behaviour will be explored further in section \ref{population}.

An  X-ray luminosity upper limit means that observed X-ray Eddington ratios $l_X \equiv L_{X, \rm max}/L_{\rm edd}$ are always smaller than unity. Furthermore, the strong suppression of thermal X-ray emission for black hole masses $M > M_{\rm edd}$ can lead to potentially large ranges of observed $l_X$ over small black hole mass ranges (figure \ref{edrats}). This has been observed by Wevers (2020), who found X-ray Eddington ratios spanning 4 orders of magnitude in a small sample of 7 X-ray TDEs.

\section{X-ray obscuration due to large mass outflows from low black hole mass TDEs}\label{obscuresec}

In addition to predicting an upper observed X-ray luminosity scale in disc-dominated TDEs, our TDE disc model makes a second prediction: that large amounts of the stellar debris will be expelled from TDEs around low-mass ($M \lesssim 10^6 M_\odot$) black holes.  Explicitly the outflow mass in our simple model is given by 
\beq
M_{\rm out} = M_{\rm deb} \, \left[ 1 - \left( {M \over M_{\rm edd}}\right)^{11/5} \right], \quad M \leq M_{\rm edd} , 
\eeq
meaning that the expelled mass very quickly becomes comparable to the surviving disc mass:
\beq
M_{\rm out} = M_d = 0.5 M_{\rm deb} \quad {\rm for} \quad M = 0.73 M_{\rm edd}. 
\eeq

\begin{figure}
  \includegraphics[width=.5\textwidth]{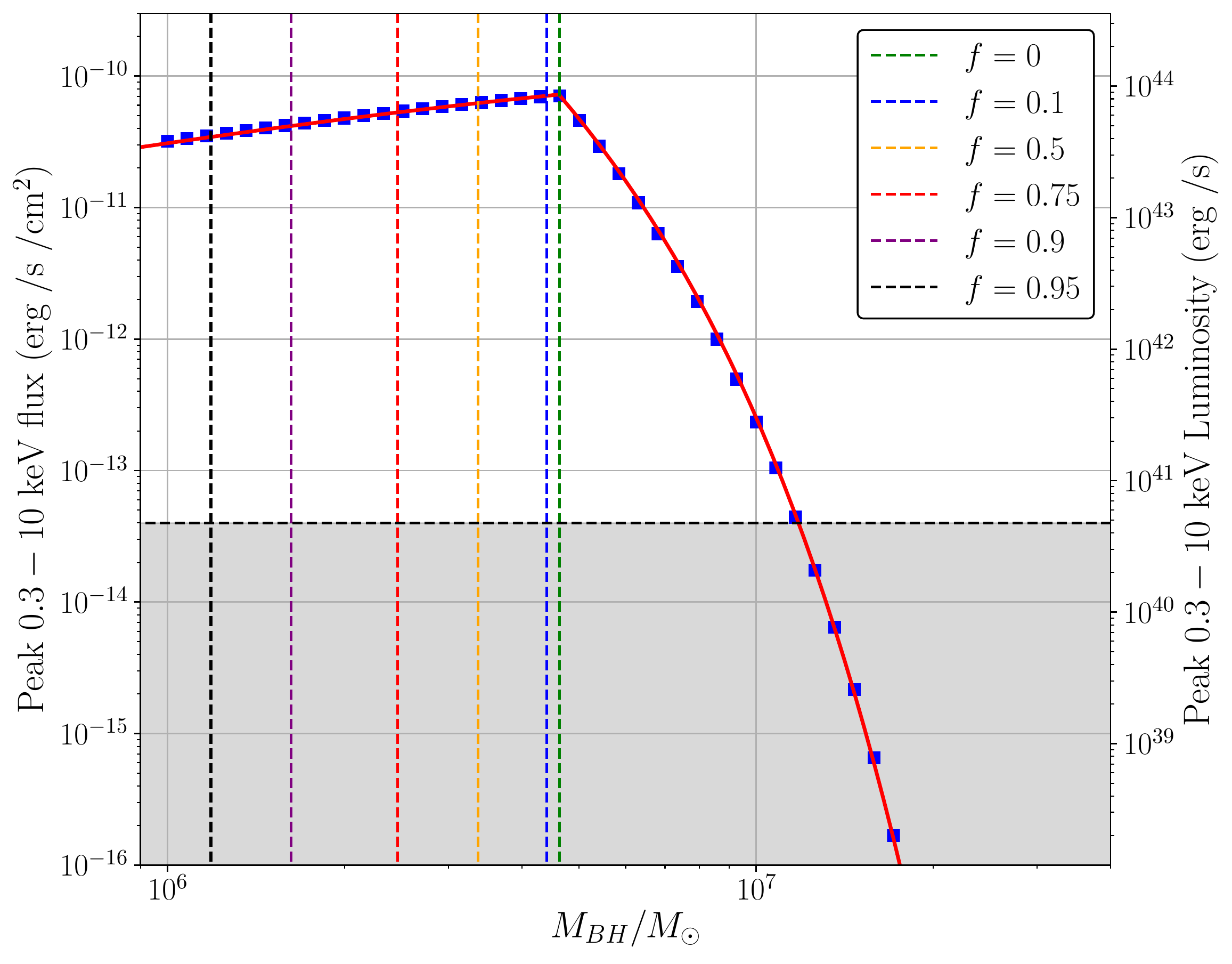} 
 \caption{The peak observed $0.3-10$ keV flux, calculated as in figure \ref{simple}, for debris mass $M_{\rm deb} = 0.5 M_\odot$ and $\alpha = 0.1$.  Over plotted as vertical dashed lines are the fraction $f$ of the initial debris mass which has been expelled from the disc system. At low black hole masses  ($M \simeq 10^6 M_\odot$) the vast majority $f > 0.95$ of the initial disc mass has been expelled. This material will likely obscure the X-ray emitting disc region from observations.   } 
 \label{simple_f}
\end{figure}

\begin{figure}
  \includegraphics[width=.5\textwidth]{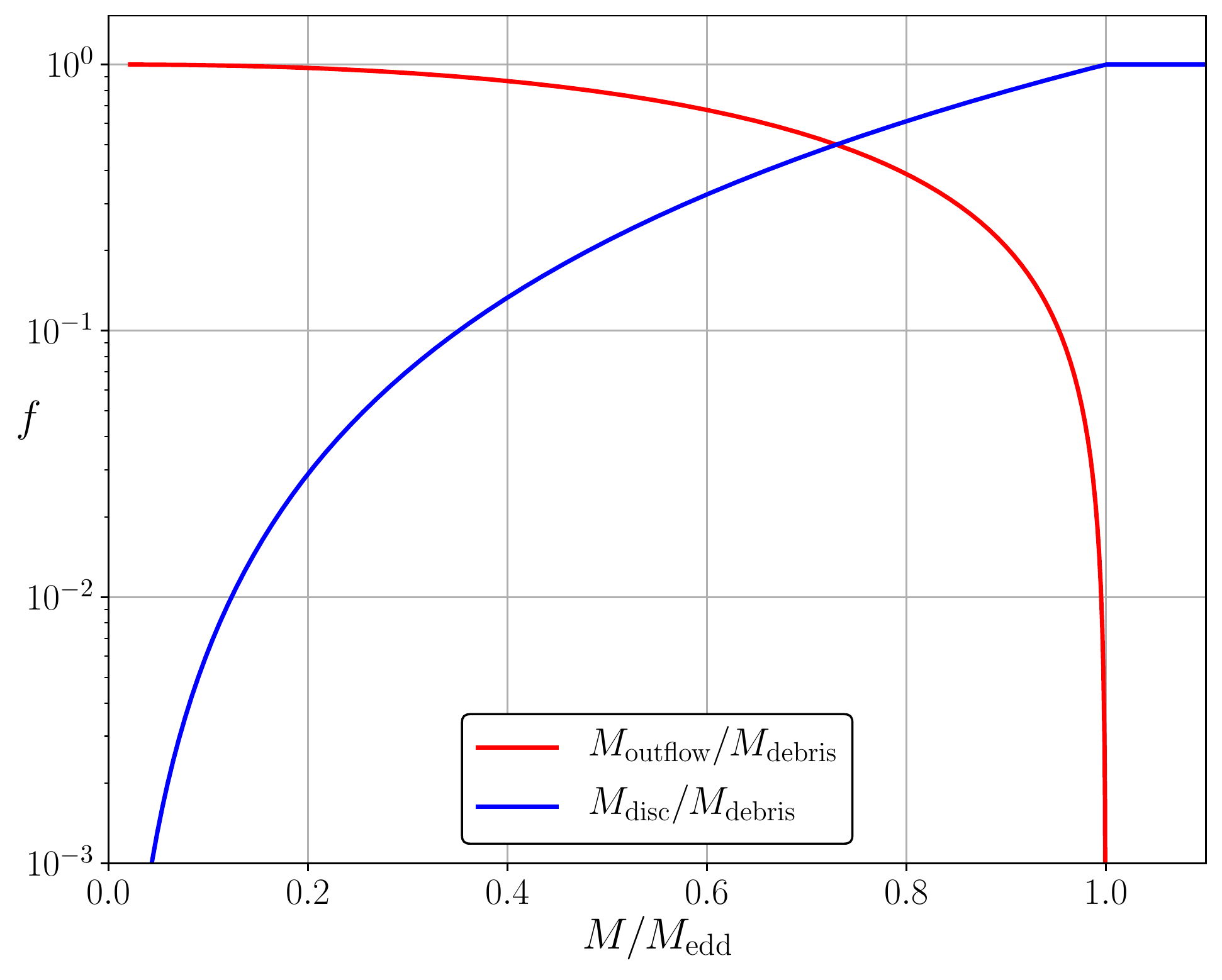} 
 \caption{The fraction of the initial debris matter which is expelled from the system (red curve), and which forms into a disc (blue curve) as a function of the black hole mass in units of $M_{\rm edd}$, the black hole mass at which all of the debris mass can form into a disc with Eddington ratio $l = 1$.    } 
 \label{fs}
\end{figure}

{In Figure \ref{simple_f}  we denote by vertical dashed lines the fraction $f$ of the initial debris mass expelled from the TDE system for the parameters used in  Figure \ref{simple}: a TDE system with an initial debris mass $M_{\rm deb} = 0.5 M_\odot$ and $\alpha = 0.1$, around Schwarzschild black holes of varying masses.} We see that TDEs around black holes with masses much smaller than $M_{\rm edd}$ result in only a tiny fraction of the debris mass forming into an Eddington-limited disc. As an explicit example,  the disc parameters $M_{\rm deb} = 0.5 M_\odot$ \& $\alpha = 0.1$ around a Schwarzschild black hole of mass $M = 10^6M_\odot$ lead to an outflow mass $M_{\rm out} = 0.484 M_\odot = 0.968 M_{\rm deb}$.  The surviving and expelled mass fractions, as a function of black hole mass, are explicitly displayed in in figure \ref{fs}. 

It is likely that for low black hole masses this large outflow of material will obscure the innermost X-ray producing regions of the accretion disc.  Metzger \& Stone (2016) studied the observed properties of TDEs in the limit where the disc formation fraction $f_{\rm in} \equiv M_d / M_{\rm deb} \ll 1$. (Notation of Metzger \& Stone 2016.) This corresponds to the $M \ll M_{\rm edd}$ limit of our model. They demonstrated that in this limit the ejected material will remain sufficiently neutral to block hard EUV and X-ray radiation from the hot inner disk, which becomes trapped in a radiation-dominated nebula. Ionising radiation from this nebula then heats the inner edge of the ejecta to temperatures of $T \sim$ few $\times 10^4$ K, converting the emission to optical/near-UV wavelengths. At these frequencies the photons more readily escape the nebula due to the lower opacity (see Metzger \& Stone (2016) for full details). Metzger \& Stone (2016) linked the launching of the radiative outflow to the rate at which the disrupted stellar material returns to the pericentre of the stars orbit ($\dot M_{\rm fb}$), which can be formally `super-Eddington' (i.e., the matter can return to pericentre at a larger rate than the formal Eddington accretion rate $\dot M_{\rm edd}$), even when the discs bolometric luminosity is sub-Eddington. In our model we link the launching of the radiative outflow directly to the discs bolometric luminosity $L$, which means this obscuring outflow is important (for typical disc parameters) for black hole masses $M \sim 10^6 M_\odot$, which is a smaller mass scale than the Metzger \& Stone (2016) model predicts. 

Our disc Eddingtonization model therefore makes a clear prediction: TDEs around low-mass black holes $M\lesssim {\rm few} \times 10^6 M_\odot$ will launch such massive outflows that the inner X-ray producing region of the disc will be obscured from observation, and these TDEs will be only observed at optical and UV frequencies. This prediction can be simply tested by inferring the black hole masses of different types of TDEs from  galactic scaling relationships (e.g. Wevers {\it et al}. 2019a).

\begin{figure}
  \includegraphics[width=.5\textwidth]{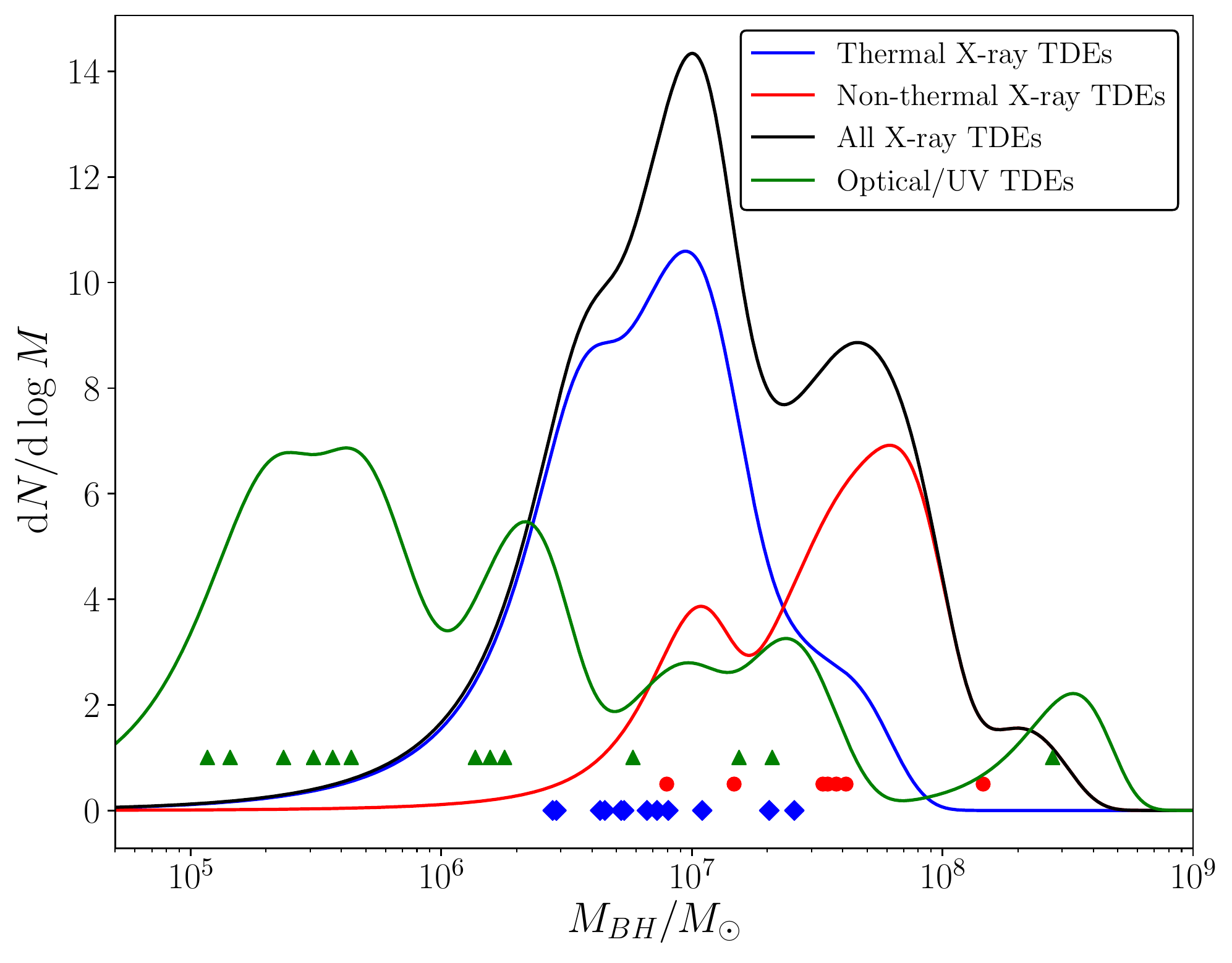} 
 \caption{The black hole masses of thermal X-ray TDEs (blue points \& curve, Paper I), non-thermal X-ray TDEs (red points \& curve,  Paper II) and optical/UV-only TDEs (green points \& curve, Wevers {\it et al}. (2019), their table A1). The inferred distributions (solid curves) are calculated  using kernel density estimation using a kernel width equal to the uncertainty in each TDEs black hole mass. The population of optical/UV-only TDEs fit exactly as predicted by the disc Eddingtonization model. All TDEs with inferred masses $M< 2\times 10^6M_\odot$ are only observed in optical and UV bands (corresponding to the 9 TDEs with the lowest inferred masses). The larger mass $M \sim 10^7 M_\odot$ optical/UV-only TDEs likely lack X-ray radiation due to the suppression of X-ray emission from large mass black holes (Paper I). } 
 \label{opt}
\end{figure}

In figure \ref{opt} we compare the inferred black hole masses of thermal X-ray TDEs (blue points \& curve, taken from Paper I), non-thermal X-ray TDEs (red points \& curve, taken from {Mummery \& Balbus 2021b; hereafter Paper II}) and optical/UV-only TDEs (green points \& curve), whose data are taken from the velocity dispersion ($\sigma$) measurements of Wevers {\it et al}. (2019a), their table A1. We include 13 of the 15 optical TDEs catalogued by Wevers {\it et al}., but do not include ASASSN-14li and ASASSN-15oi as they were also observed at X-ray energies. We calculate the black hole masses of the optical/UV TDEs from the $M:\sigma$ relationship of McConnell \& Ma (2013). The inferred distributions (solid curves) are calculated using kernel density estimation with a kernel width equal to the uncertainty in each TDEs black hole mass.

It is clear from figure \ref{opt} that the population of optical/UV-only TDEs fit exactly as predicted by this Eddingtonization model. All TDEs with inferred masses $M< 2\times 10^6M_\odot$ were only observed in optical and UV bands (corresponding to the 9 TDEs with the lowest inferred masses). The larger mass $M \sim 10^7 M_\odot$ optical/UV-only TDEs likely lack X-ray radiation due to the suppression of X-ray emission from large mass black holes (Paper I). The extremely massive outlier is ASASSN-15lh, which we have studied in detail in a previous work (Mummery \& Balbus 2020b). Finally, we note that the $M:\sigma$ relationship of McConnell \& Ma (2013) is not well calibrated for masses $M < 10^6 M_\odot$, and so while the lowest mass TDE hosts in figure \ref{opt} certainly have small black hole masses, the exact values inferred from the scaling relationship should be treated as having a larger than normal uncertainty associated with them. 

\section{Maximum X-ray luminosity of hard state X-ray TDEs}\label{hardnum}

In Paper II we developed a model of hard X-ray emission resulting from TDE systems with Eddington ratios $l \lesssim 10^{-2}$. {This hard state emission results} from the Compton up-scattering of thermal disc photons by an electron scattering corona. This model was parameterised by a coronal radius $R_{\rm Cor}$ (the outer radius of the corona), the fraction of photons scattered by the corona $f_{SC}$, and a photon index $\Gamma$. {(The same parameters as the SIMPL model (Steiner et al. 2009) in XSPEC (Arnaud 1996))}. The brightest hard X-ray sources will result from coronae which scatter all of the soft disc photons ($f_{SC} \rightarrow 1$).  While this limit is somewhat unphysical, it allows us to determine an upper luminosity scale for hard X-ray TDE sources, as well as for thermal TDEs (figures \ref{spins}, \ref{VSspins}). 

\begin{figure}
  \includegraphics[width=.5\textwidth]{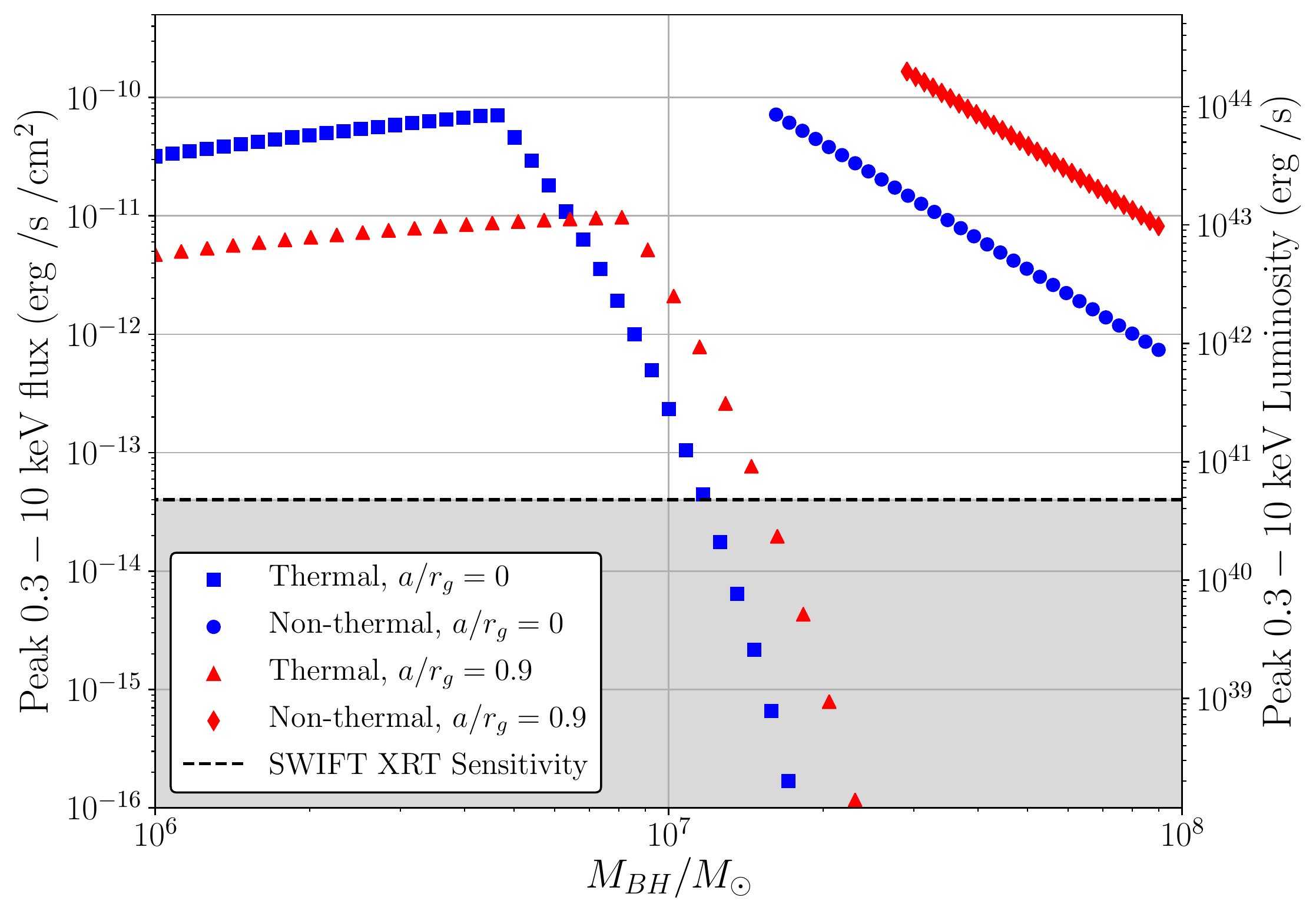} 
 \caption{The peak observed $0.3-10$ keV flux, as observed from face-on accretion discs at a distance $D = 100$ Mpc, for black hole masses between $10^6$ \& $10^8 M_\odot$. Blue points are for a Schwarzschild black hole, and red points are for a more rapidly rotating Kerr black hole ($a/r_g = 0.9$). In the thermal emission regime the fluxes are calculated as in figure \ref{simple}. For sufficiently large black hole masses (when the discs Eddington ratio is less than $10^{-2}$) non-thermal X-ray emission is calculated by the model of Paper II, assuming all photons in the inner disc regions ($r < 12r_g$) are Compton scattered ($f_{SC} = 1$) with a photon index $\Gamma = 2$. Despite having entirely different emission mechanisms the maximum X-ray luminosities in the two regimes are comparable.   } 
 \label{hard}
\end{figure}

In Figure \ref{hard} we numerically calculate the thermal and non-thermal emission from TDE systems with either a Schwarzschild or a more rapidly rotating ($a/r_g = 0.9$) Kerr black hole. These events had a debris mass $M_{\rm deb} = 0.5 M_\odot$, an $\alpha$-parameter $\alpha = 0.1$, and were observed face-on at a distance of $D = 100$ Mpc.  If the bolometric {luminosity} of the resulting disc is greater than the Eddington luminosity then the disc mass is reduced according to equation \ref{survive}. Conversely, if the Eddington ratio is lower than $l = 0.01$ we modelled the emission with the compact corona model of Paper II, with $f_{SC} = 1$, $\Gamma = 2.0$ and $R_{\rm Cor} = 12 r_g$. 

A very interesting result highlighted by figure \ref{hard} is that the upper X-ray luminosity scale of TDE disc systems in the non-thermal state is very similar to the maximum X-ray luminosity scale in the thermal state (figs. \ref{spins}, \ref{VSspins}). This is despite the two regimes involving very different physical emission models. These results therefore suggest that, irrespective of black hole mass and accretion state, TDEs observed at X-ray energies will always be observed at luminosity scales less than the maximum luminosity derived here
\beq
L_{X, \rm peak} < L_M \simeq  10^{44} \,\, {\rm erg/s}.
\eeq
In the following section we analyse the peak observed X-ray luminosities of the historical X-ray TDE population. 

\section{X-ray TDE population analysis}\label{population}

To date the author is aware of 24 TDEs which have been well observed at X-ray energies at times near to the peak of the tidal disruption flare (in this section we ignore jetted TDEs). Within this sample of TDEs there is a (potentially) surprising variety in the observed properties of individual TDE X-ray light curves. We argue here that, despite the apparent variety in observed properties, the historic population of X-ray TDEs do in fact have a unifying property: the amplitude of the peaks of their X-ray light curves correspond to the limiting luminosities derived in this paper.

The X-ray light curves of well observed TDE flares can be loosely split into 3 different types: `well behaved' light curves which smoothly decay with time; flaring TDEs which show order of magnitude fluctuations over timescales as short as days to weeks; and late time re-brightening TDEs, who's X-ray luminosity grows by orders of magnitude hundreds of days after the initial flare. Some example light curves of each type are shown below.

The fact that individual TDE light curves show such varied behaviour, except for the amplitude of their peak X-ray luminosities, suggests that the X-ray luminosity limit derived in this paper is a universal feature of these events.
\begin{figure}
  \includegraphics[width=.5\textwidth]{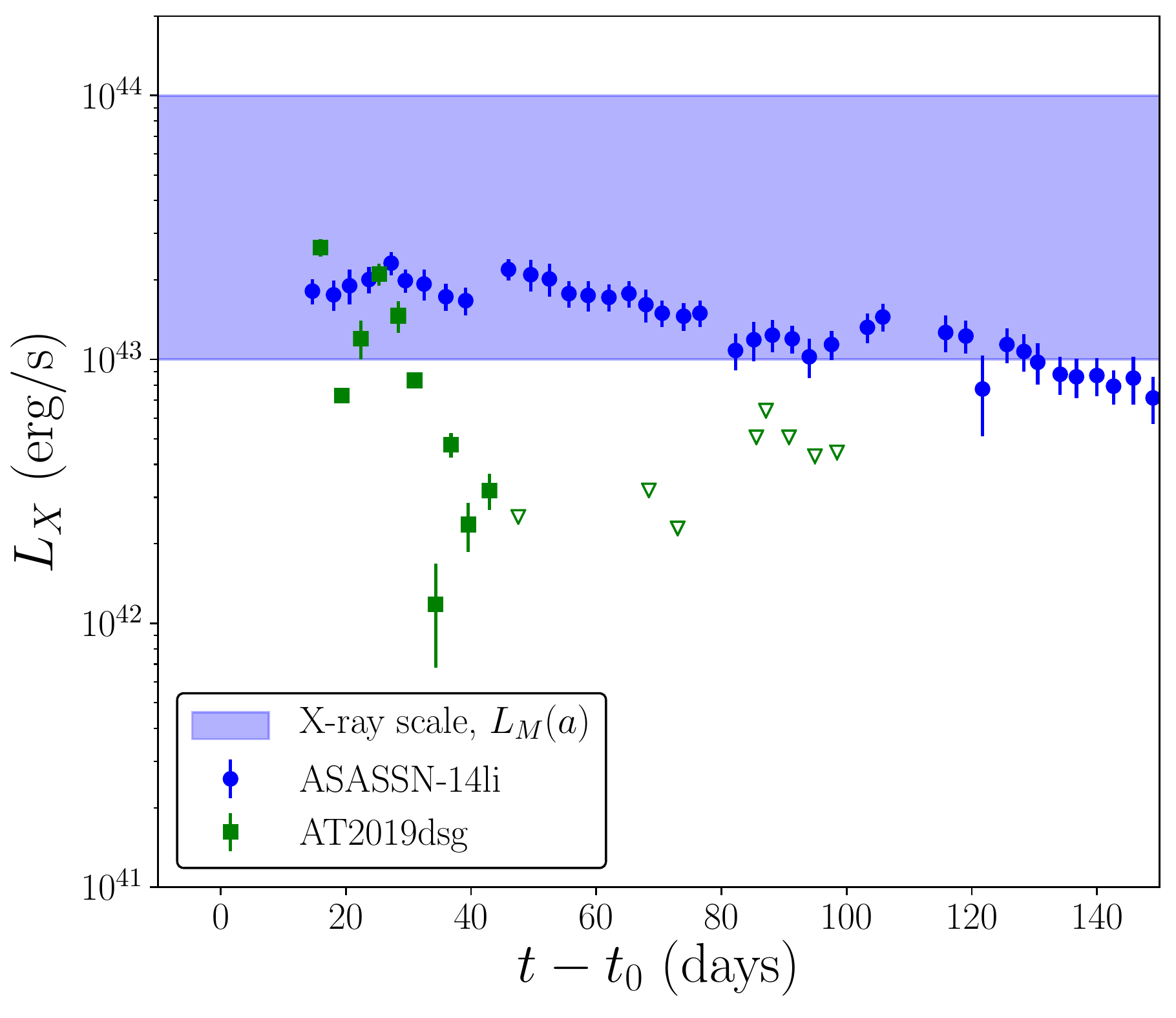} 
 \caption{The observed light curves of two `well-behaved' X-ray TDEs ASASSN-14li and AT2019dsg. The blue shaded region corresponds to the characteristic luminosity scale, a function of black hole spin, derived in this paper. Despite having very different decay timescales, both TDEs peak in the anticipated luminosity region. Upper detection limits of AT2019dsg are displayed by inverted triangles. } 
\label{compsmooth}
\end{figure}
\begin{figure}
  \includegraphics[width=.5\textwidth]{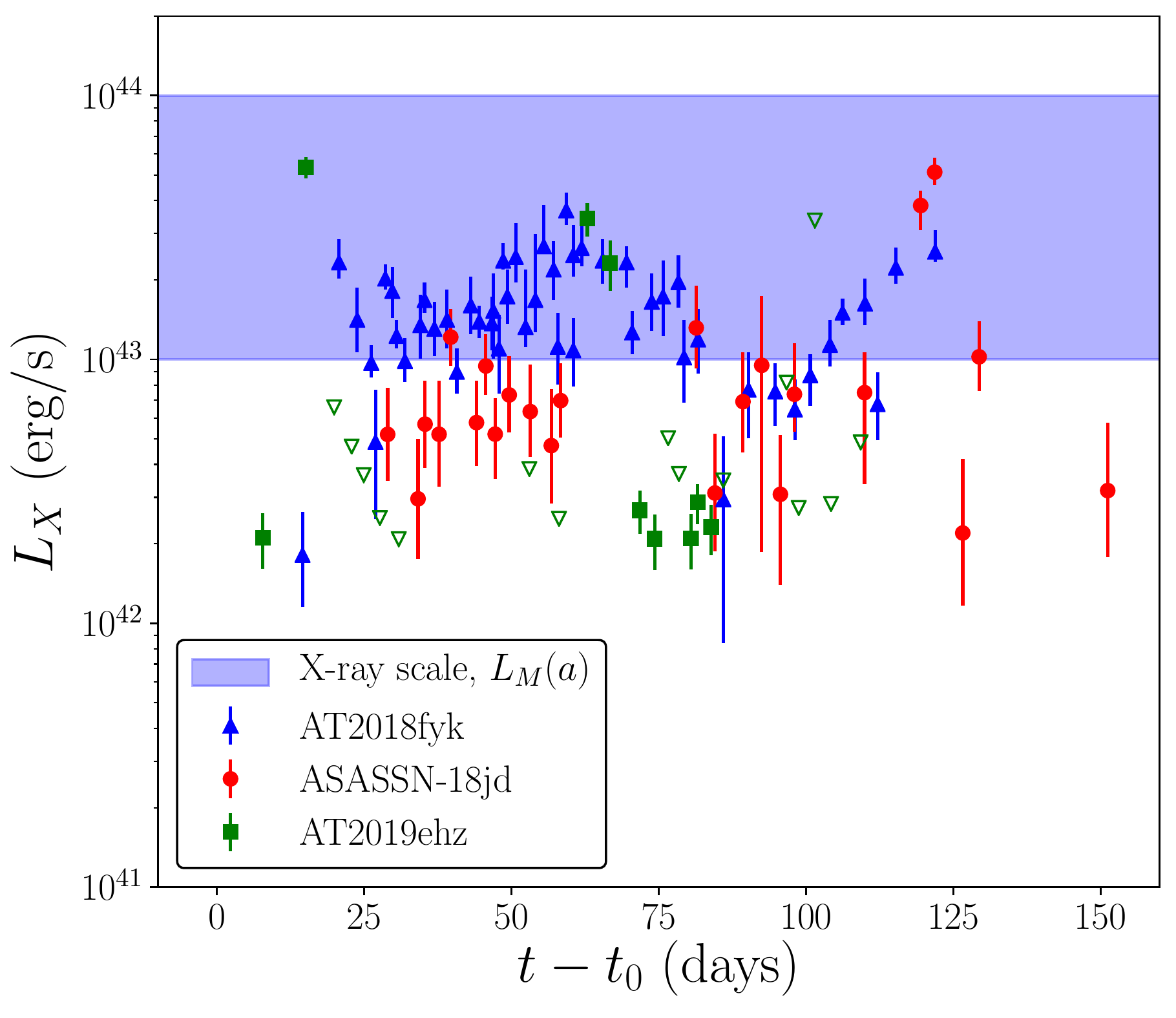} 
 \caption{The observed light curves of three X-ray TDEs (AT2018fyk, ASASSN-18jd and AT2019ehz) which display extreme short timescale variability. The blue shaded region corresponds to the characteristic luminosity scale, a function of black hole spin, derived in this paper. Despite having extremely complex temporal properties, all three TDEs peak in the anticipated luminosity region. Upper detection limits of AT2019ehz are displayed by inverted triangles.} 
 \label{compflare}
\end{figure}
\begin{figure}
  \includegraphics[width=.5\textwidth]{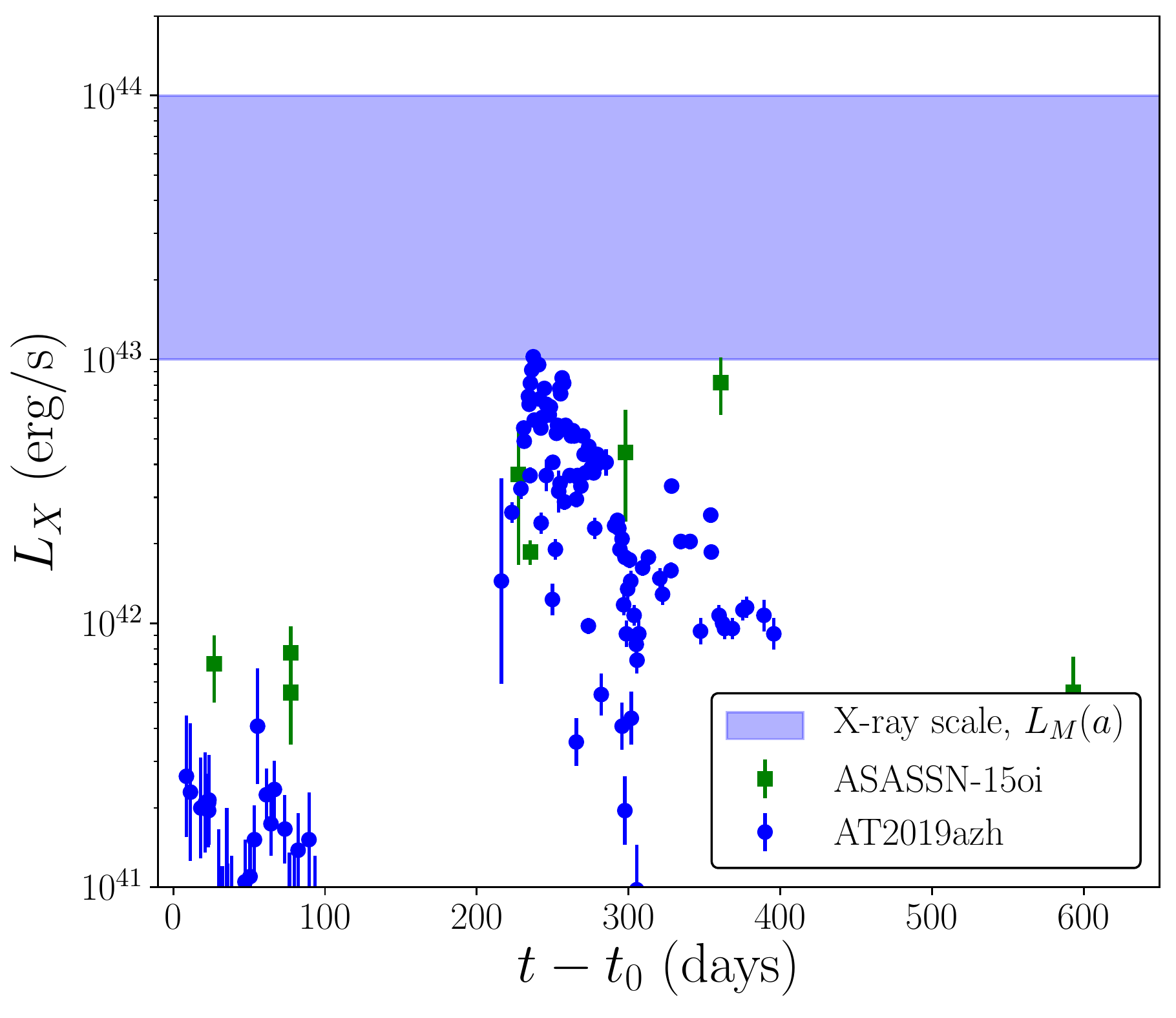} 
 \caption{The observed light curves of two X-ray TDEs (ASASSN-15oi and AT2019azh) which re-brighten by an order of magnitude at late times. The blue shaded region corresponds to the characteristic luminosity scale, a function of black hole spin, derived in this paper. Although initially X-ray dim, after re-brightening these TDEs peak with luminosities at the characteristic luminosity scale. } 
 \label{complate}
\end{figure}

\subsection{`Well behaved' X-ray TDEs}

`Well-behaved' TDEs are those which have been observed to quickly form X-ray bright accretion discs in the aftermath of the stellar disruption, which then monotonically fade with time. Perhaps the most well known example is ASASSN-14li, which underwent an X-ray flare of peak magnitude $L_X \simeq 2 \times 10^{43}$ erg/s, before fading over the next $1000$ days (Holoein {\it et al}. 2016a, Bright {\it et al}. 2018, Mummery \& Balbus 2020a). The first 150 days of ASASSN-14li's X-ray light curve are shown as blue dots in Fig. \ref{compsmooth}. A more recently discovered `well-behaved' X-ray TDE is AT2019dsg (van Velzen {\it et al.} 2020, green squares). AT2019dsg  had a peak X-ray luminosity of $L_X \simeq 3 \times 10^{43}$ erg/s, before rapidly dimming and becoming unobservable within 60 days of the first observation. The blue band in figure \ref{compsmooth} corresponds to the typical X-ray luminosity scale (a function of black hole spin) derived in this paper. 

The very different evolutionary time scales of these two TDEs likely means that either some aspect of the disc microphysics (i.e., the $\alpha$ parameter or disc mass), or the black hole mass of the two sources are very different. However, both sources have peak X-ray luminosities exactly in line with the theoretical calculations presented in this paper. 

\subsection{Flaring TDEs}

Some X-ray TDEs show extreme X-ray variability on timescales as short as days to weeks.  Three example of such events are shown in Figure \ref{compflare}, where we plot the observed X-ray light curves of AT2018fyk (Wevers {\it et al}. 2019, blue triangles), ASASSN-18jd (Neustadt {\it et al}. 2020, red dots) and AT2019ehz (van Velzen {\it et al}. 2020, green squares) and our theoretical X-ray limiting region  (blue shaded region). We observe that all three light curves, despite their extremely complex temporal behaviour, peak in the limiting luminosity range. 

The ultimate cause of the large amplitude fluctuations in the X-ray luminosity are not at present clear: they could result from intrinsic variations in accretion rate, or by the fallback of matter scattered onto large radius orbits in the initial disruption onto the disc, or even time varying obscuration of the inner disc by other TDE debris. Irrespective of cause,  the magnitude of the fluctuations are limited to the same luminosity scale as the more stably evolving discs which produce the light curves of `well behaved' TDEs like ASASSN-14li. 

\subsection{Late time re-brightening TDEs}
Finally, other X-ray TDEs have been observed which appear to be X-ray dim at early times, but then undergo a re-brightening by as much as two orders of magnitude at much later times. Examples of this type of X-ray TDE are ASASSN-15oi (Holoein {\it et al}. 2018) and AT2019azh (van Velzen {\it et al}. 2020, Hinkle {\it et al}. 2021), their light curves are plotted in Figure \ref{complate}.  

The physics of disc formation in TDEs which are initially X-ray dim may be very different from that of the TDEs which are X-ray bright at early times. It is therefore encouraging that when they do eventually re-brighten, the peak luminosities are again consistent with those expected of Eddington limited accretion discs.

\subsection{The total X-ray TDE population}
\begin{figure}
  \includegraphics[width=.5\textwidth]{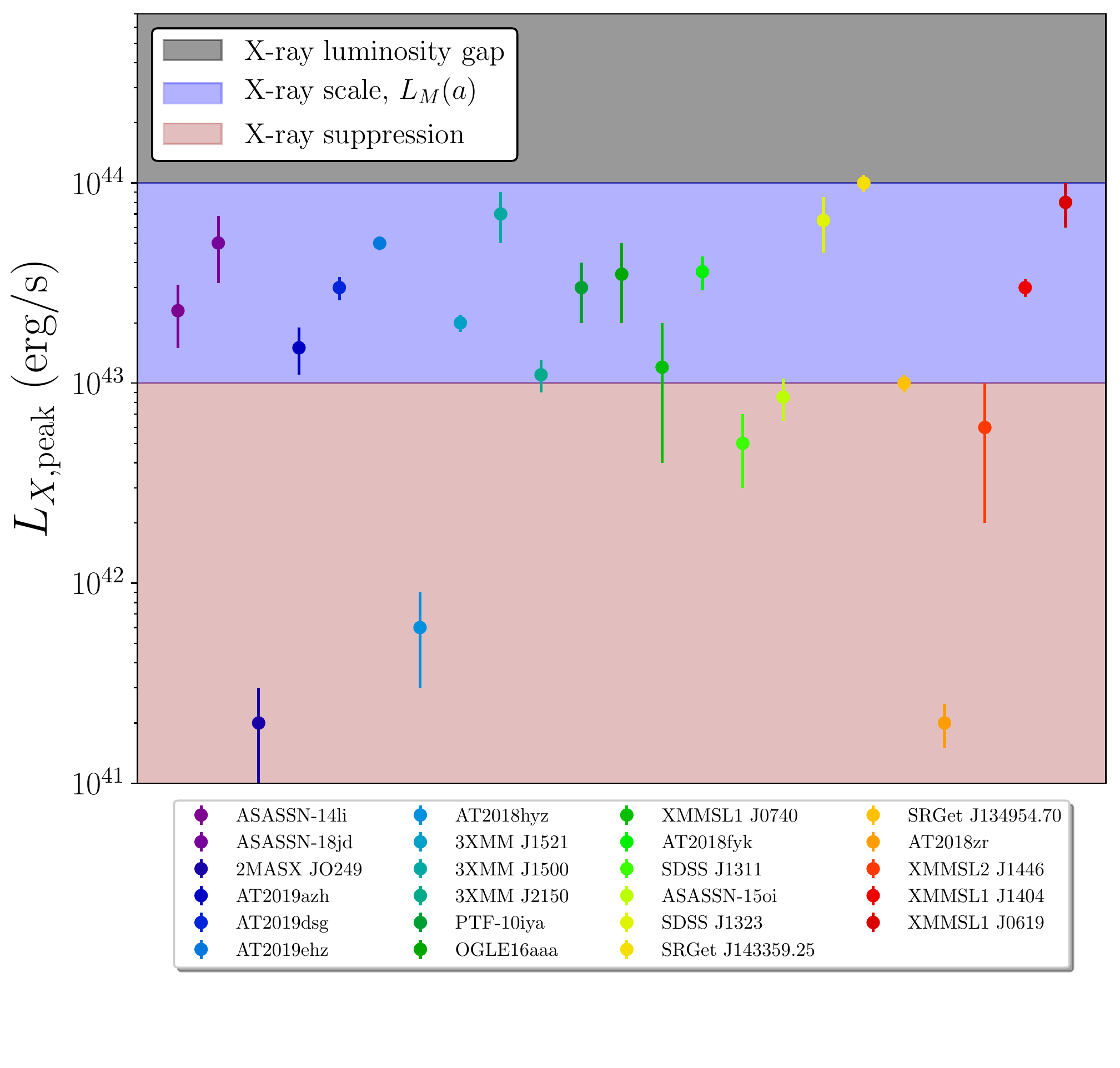} 
 \caption{The peak observed X-ray ($0.3$--$10$ keV) luminosities of the historic TDE population (data in table \ref{comptable}). 20 of the 24 disc-dominated X-ray TDEs lie at the black hole spin-dependent  characteristic luminosity scale derived in this paper.  No disc dominated TDEs have yet been observed in the predicted barren region of the X-ray TDE luminosity diagram, with $L_X > 10^{44}$ erg/s. } 
 \label{popsoft}
\end{figure}

In figure \ref{popsoft} we plot the peak observed X-ray luminosities of the 24 disc-dominated X-ray TDEs observed to date. References and exact luminosity values can be found in Table \ref{comptable} in Appendix \ref{dat}.  A final TDE NGC 247 peaked at $L_X = 2 \times10^{39}$ erg/s and is not displayed for aesthetic reasons, but is included in Table \ref{comptable}. 

\begin{figure}
  \includegraphics[width=.5\textwidth]{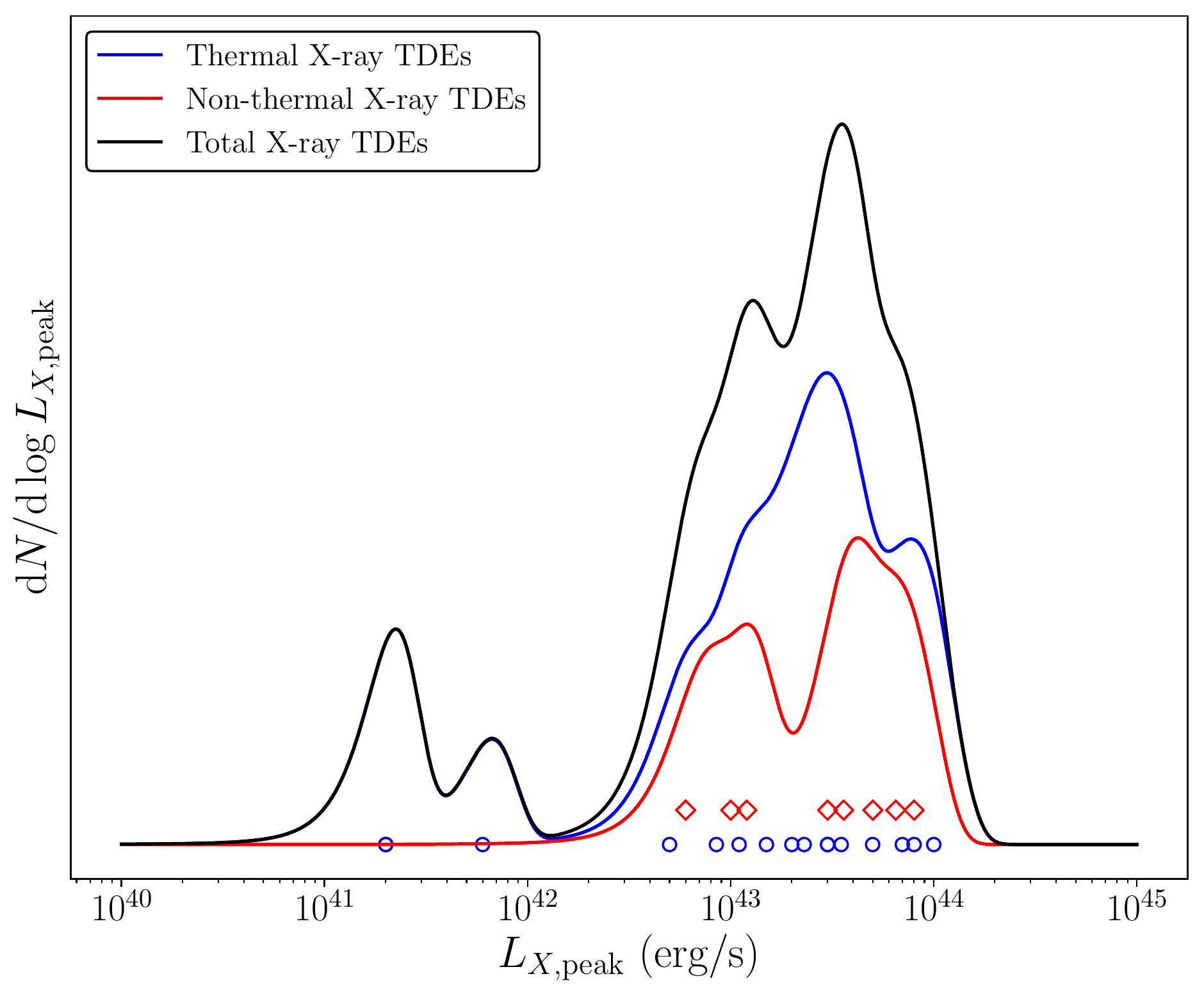} 
 \caption{The peak observed X-ray ($0.3$--$10$ keV) luminosities of the historic TDE population (data in table \ref{comptable}). 20 of the 24 disc-dominated X-ray TDEs lie in the black hole spin-dependent  characteristic luminosity scale derived in this paper.  No disc dominated TDEs have yet been observed in the predicted barren region of the X-ray TDE luminosity diagram, with $L_X > 10^{44}$ erg/s. } 
 \label{lumdist}
\end{figure}

 The black-shaded region of the plot, defined by X-ray luminosities  $L_X > 1 \times 10^{44} \,{\rm erg/s} $ is the anticipated `barren region' of the observed TDE X-ray luminosity diagram. This region of X-ray luminosity parameter space is higher than the peak X-ray luminosities of both the thermal disc solutions in the Eddingtonization regime and the nonthermal Compton-scattered solutions (figure \ref{hard}), but is below the expected X-ray luminosity scale of jetted TDEs. Very few X-ray bright TDEs are therefore expected to be observed at this luminosity scale. Of the 24 TDEs detected to date not one has been observed at this luminosity scale. 
 
 The blue shaded region, $10^{43} < L_X ({\rm erg/s}) < 10^{44}$, corresponds to the theoretical X-ray luminosity scale, a function of black hole spin, derived in this paper (figure \ref{spins}, \ref{VSspins}, \ref{hard}). 20 of the 24 X-ray TDEs peak within this region, despite having light curves with otherwise widely varying properties. This is an extremely telling result. The inferred distribution of the X-ray luminosity of quasi-thermal (blue dots) \& non-thermal (red diamonds) X-ray TDEs is displayed in figure \ref{lumdist}. The inferred distributions (solid curves) are calculated using kernel density estimation with a kernel width equal to the uncertainty in each TDEs peak X-ray luminosity. X-ray TDEs of both spectral types peak at the same luminosity scale. 
 
\begin{figure}
  \includegraphics[width=.5\textwidth]{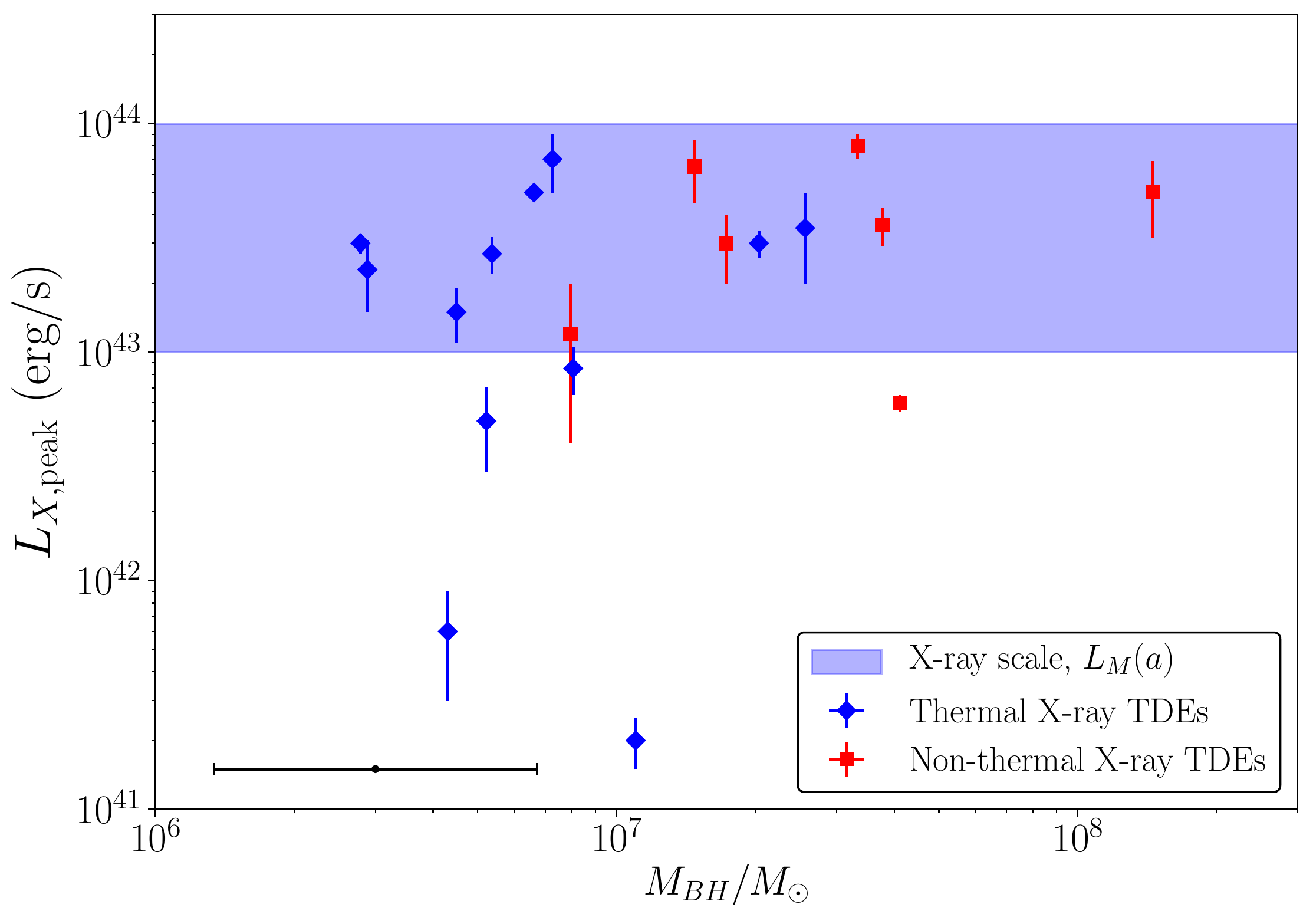} 
 \caption{The peak observed X-ray luminosities of the 19 X-ray TDEs with black hole mass estimates from galactic scaling relationships (Papers I \& II). Different X-ray spectral types are denoted by blue diamonds (quasi-thermal emission) and red squares (non-thermal emission).  The black point shows the typical uncertainty in the black hole masses resulting from intrinsic scatter in galactic scaling relationships. While there is a systematic difference in the black hole masses of the different X-ray spectral types, the peak X-ray luminosity is approximately independent of the central black hole mass.  } 
 \label{MLscat}
\end{figure}

 Finally, we can rule out the naive model that there exists some intrinsic relationship between the Eddington and X-ray luminosities of X-ray bright TDEs (i.e., that the X-ray luminosity is some fixed fraction of the Eddington luminosity). If the X-ray luminosity of TDEs were in some way related to the black holes Eddington luminosity, then observed X-ray luminosity amplitudes should vary between TDEs around black holes of different masses:
  \beq
 L_{\rm edd} = 1 \times 10^{45} \left({M \over 8 \times 10^6 M_{\odot}}\right) \, {\rm erg/s} .
 \eeq
In figure \ref{MLscat} we plot the peak observed X-ray luminosity of the 19 X-ray TDEs which also have black hole mass estimates from galactic scaling relationships (see Papers I \& II for a summary of the black hole mass measurements), the black point in figure \ref{MLscat} shows a typical uncertainty on the black hole masses resulting from intrinsic scatter in galactic scaling relationships.   Different X-ray spectral types are denoted by blue diamonds (quasi-thermal emission) and red squares (non-thermal emission).  
 
 While the black hole masses of the different spectral types of X-ray TDEs are systematically different (Paper II), figure \ref{MLscat} demonstrates that the peak X-ray luminosity of TDE sources is in fact effectively independent of black hole mass.  This observation is strongly supportive of the disc models put forward in this paper.

 We believe that the strong agreement between the observed properties of the X-ray TDE population and the predictions of our TDE disc model (figures \ref{opt}, \ref{lumdist}, \ref{MLscat}) provide a deeper understanding of the underlying TDE evolution and compelling evidence for disc-dominated TDEs being the norm.

{\section{Conclusions}\label{conc}
\begin{figure}
  \includegraphics[width=.5\textwidth]{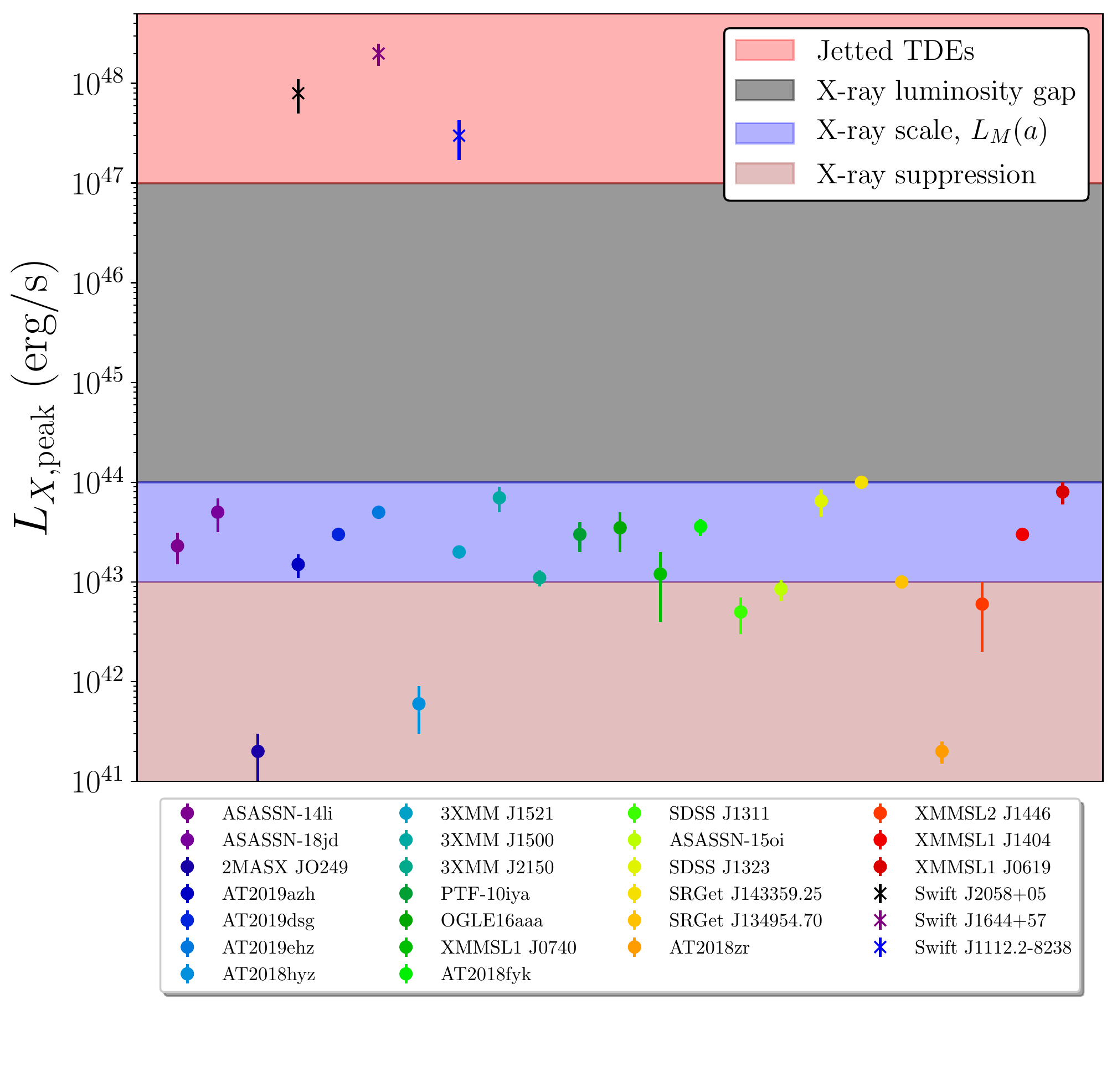} 
 \caption{The peak observed X-ray ($0.3$--$10$ keV) luminosities of the historic TDE population, disc-dominated X-ray sources are denoted with dots (data in table \ref{comptable}) and jetted TDEs are denoted with crosses. Also plotted are four shaded regions which denote the different luminosity scales derived in this paper. 20 of the 24 soft X-ray TDEs lie at the characteristic luminosity scale derived in this paper. No disc dominated TDEs have yet been observed in the predicted barren region of the X-ray TDE luminosity diagram, with $10^{47} > L_X > 10^{44}$ erg/s.  } 
 \label{poptot}
\end{figure}

In this paper we have built upon and extended existing models of outflows launched from TDE discs at early times. In particular, in contrast with previous approaches, we have coupled the launching of a radiatively driven outflow directly to the bolometric luminosity of the TDE accretion disc (in Eddington units).  A lack of observed TDEs in super-Eddington accretion states (e.g. Jonker {\it et al}. 2020) can be naturally explained if TDE discs launch outflows at the earliest times in their evolution, thereby limiting the amount of the stellar debris which can form into an accretion disc in close proximity to the black hole.  In this paper we have assumed that the amount of material which can form into an accretion disc in the aftermath of a TDE is set such that the discs resulting luminosity is limited by the black holes Eddington luminosity. 

This simple assumption about TDE disc formation leads to a number of quantitative and testable predictions about the properties of the populations of X-ray bright TDEs and optically bright X-ray dim TDEs. Our main predictions are the following. Firstly, there should exist a population of TDEs observed only at optical and UV frequencies which occur around black holes of the lowest masses $M \lesssim {\rm few} \times 10^6 M_\odot$. Physically this results from the strong inverse dependence of TDE disc Eddington ratio on central black hole mass (eq. \ref{edrat}), meaning that TDEs around the lowest mass black holes will result in extreme initial luminosities. The resulting large-mass outflow (fig. \ref{fs}) will obscure the inner-most disc regions from observers, reprocessing the high energy inner disc emission to optical emission. The existing population of optical/UV-only TDEs fit exactly as predicted by this Eddingtonization model (fig. \ref{opt}). All TDEs with inferred masses $M< 2\times 10^6M_\odot$ were only observed in optical and UV bands (corresponding to the 9 known TDEs with the lowest inferred masses).

Our second key prediction concerns the properties of the observed amplitudes of the X-ray luminosity of the population of X-ray bright TDEs. We demonstrate analytically and numerically (eq. \ref{fluxM}, figs. \ref{simple}, \ref{spins}, \ref{VSspins}) that the X-ray luminosity of TDE discs with bolometric luminosities comparable to their Eddington luminosity are approximately constant (i.e. independent of black hole mass) across the black hole mass range of interest for TDEs. The amplitude of this X-ray luminosity is  $L_M \simeq 10^{43}-10^{44}$ erg/s, where the primary variance results from the range of allowed black hole spins. A similar limiting X-ray luminosity exists for TDE discs evolving in harder accretion states, with additional nonthermal emission resulting from a Compton scattering corona (fig. \ref{hard}), and so this limiting X-ray luminosity should be imprinted on the X-ray luminosity distribution of observed TDEs. 

The properties of the existing population of TDEs are in excellent accord with this prediction (figs. \ref{popsoft}, \ref{lumdist}), with 20 of 24 TDEs having peak X-ray luminosities which lie exactly at the scale derived in this work. The remaining 4 TDEs all had peak X-ray luminosities lying below the limiting scale, meaning that no disc-dominated TDE has been observed with X-ray luminosity above our predicted upper limit. This is despite the observed light curves of these 24 X-ray bright TDEs having otherwise widely varying temporal properties (figs. \ref{compsmooth}, \ref{compflare}, \ref{complate}). TDEs with X-ray luminosity resulting from a powerful radio jet can comfortably exceed $L_M \sim 10^{44}$ erg/s, but normally by many orders of magnitude (fig. \ref{poptot}). An X-ray luminosity gap spanning roughly 3 order of magnitude $L_X \sim 10^{44} - 10^{47}$ erg/s, where no TDE X-ray light curves peak, is  a natural prediction of the model put forward in this paper. 

The synthesis of time-dependent accretion disc models (Mummery \& Balbus 2020a, b) with models of outflows from TDE discs (Metzger \& Stone 2016) results in a series of predictions which are in excellent agreement with the properties of the existing TDE population.  We believe that this work paves the way for a more concrete understanding of a wide range of observed TDE properties, whereby the Eddington ratio of the disc which forms in the aftermath of a TDE is the primary parameter which controls the observed properties of a given TDE.  As this parameter is strongly correlated with the TDEs black hole mass (eq. \ref{edrat}), a quantity which can be directly inferred from observations of the TDEs host galaxy,  a unification scheme of this form can and will be directly tested in the near future as the number of observed TDEs grow in number. 

}

\section*{Data accessibility statement}
All data used in this paper is presented in full in Appendix \ref{dat}. 
\section*{Acknowledgments}
It is a pleasure to acknowledge useful conversations with Steven Balbus and Jeremy Goodman.  The author is grateful to Lauren Rhodes for comments on an earlier version of this manuscript. We thank the anonymous referee for helpful comments.

\appendix{}
\section{Observed TDE catalogue}\label{dat}

\begin{table}
\renewcommand{\arraystretch}{2}
\centering
\begin{tabular}{|p{2.2cm}|p{2.1cm}| p{2.75 cm} |}
\hline
TDE   & $L_{X, {\rm peak}}\, $ (erg/s)  &  Reference \\ \hline\hline
ASASSN-14li & $2.3 \pm 0.8 \times10^{43}$ &  Bright {\it et al}. 2018 \\ \hline 
ASASSN-15oi & $8.5 \pm 0.1 \times10^{42}$  & Holoein {\it et al}. 2018 \\ \hline 
ASASSN-18jd & $5.1 \pm 0.5 \times10^{43}$   & Neustadt  {\it et al}. 2020  \\ \hline
AT2018zr &$2 \pm 0.3 \times10^{41}$ & van Velzen {\it et al}. 2019b \\ \hline
AT2018fyk & $4 \pm 0.5 \times10^{43}$  & Wevers {\it et al}. 2019  \\ \hline
AT2018hyzj & $6 \pm 3 \times10^{41}$  & van Velzen {\it et al}. 2020 \\ \hline
AT2019dsg & $3 \pm 0.1 \times10^{43}$  & van Velzen {\it et al}. 2020 \\ \hline
AT2019ehz &$5 \pm 0.2 \times10^{43}$ & van Velzen {\it et al}. 2020 \\ \hline
AT2019azh &$1.6 \pm 0.1 \times10^{43}$   & van Velzen {\it et al}. 2020  \\ \hline 
2MASX JO249 & $3 \pm 0.5 \times10^{41}$&Auchettl {\it et al}. 2017 \\ \hline
3XMM  J1521 &  $2.7 \pm 0.5 \times10^{43}$ &Auchettl {\it et al}. 2017 \\ \hline
3XMM  J1500 &  $7 \pm 0.7 \times10^{43}$ & Lin {\it et al}. 2017 \\ \hline
3XMM  J2150 &  $1.1 \pm 0.4 \times10^{43}$ & Lin {\it et al}. 2018 \\ \hline
NGC 247 & $2\pm0.1\times10^{39}$ &Auchettl {\it et al}. 2017 \\ \hline
OGLE16aaa & $2.2 \pm 0.4 \times10^{43}$ &Auchettl {\it et al}. 2017 \\ \hline
PTF-10iya&  $2.5 \pm 0.3 \times10^{43}$  &Auchettl {\it et al}. 2017 \\ \hline
SDSS J1311& $7 \pm 2 \times10^{42}$ &Auchettel {\it et al}. 2017 \\ \hline
SDSS J1323&$7 \pm 2 \times10^{43}$&Esquej {\it et al}. 2008 \\ \hline
XMMSL1 J0740 &$1.2 \pm 1 \times10^{43}$&Auchettl {\it et al}. 2017 \\ \hline
XMMSL2 J1446 &$6 \pm 4 \times10^{42}$&Saxton {\it et al}. 2019 \\ \hline
XMMSL1 J1404 &$3 \pm 0.3 \times10^{43}$&Wevers 2020 \\ \hline
XMMSL1 J0619 &$8 \pm 2 \times10^{43}$&Saxton {\it et al}. 2014 \\ \hline
SRGet J143359.25& $1 \times 10^{44}$ & Khabibullin {\it et al}. 2020a \\ \hline 
SRGet J134954.70 & $1 \times 10^{43}$ & Khabibullin {\it et al}. 2020b \\ \hline
\end{tabular}
\caption{The peak observed X-ray luminosity of 24 TDEs from the literature. }
\label{comptable}
\end{table}

We have collated the maximum observed X-ray luminosities of the existing X-ray TDE population. Each TDEs peak luminosity is presented in Table \ref{comptable}. The reference for each TDE corresponds to the paper from which the maximum X-ray luminosity was taken,  not necessarily the paper in which the TDE was first discovered.

\label{lastpage}
\end{document}